\documentclass[extra,mreferee]{gji2}
\usepackage{amsmath}
\usepackage{amsfonts}
\usepackage{timet}
\usepackage{url}
\usepackage{bm}
\usepackage{color}
\usepackage{graphicx}
\usepackage{multirow}
\usepackage{newfloat,caption}

\DeclareFloatingEnvironment[fileext=lop]{contfigure}
\newcommand{\lp}{\left(}
\newcommand{\rp}{\right)}
\newcommand{\llp}{\left\{}
\newcommand{\rrp}{\right\}}
\newcommand{\lllp}{\left[}
\newcommand{\rrrp}{\right]}

\newcommand{\bE}{\mathbb{E}}
\newcommand{\mD}{\mathcal{D}}
\newcommand{\mM}{\mathcal{M}}
\newcommand{\mK}{\mathcal{K}}
\newcommand{\mN}{\mathcal{N}}

\newcommand{\prior}{\mathrm{prior}}
\newcommand{\trunc}{\mathrm{trunc}}

\newtheorem{lemma}{Lemma}

\title[Forecasting aftershocks with GPR]
  {Forecasting temporal variation of aftershocks immediately after a main shock using Gaussian process regression }
\author[K. Morikawa et al.]
  {Kosuke Morikawa$^1$, Hiromichi Nagao$^{2,3}$, Shin-ichi Ito$^{2,3}$,   Yoshikazu Terada$^{1,4}$, \and Shin'ichi Sakai$^{2,5}$ and Naoshi Hirata$^{2,6}$. \\
  $^1$ Graduate School of Engineering Science, Osaka University, 1-3, Machikaneyama-cho, Toyonaka-shi, \\Osaka \emph{560-8531}, Japan.\\
  $^2$ Earthquake Research Institute, The University of Tokyo, 1-1-1, Yayoi, Bunkyo-ku, Tokyo \emph{113-0032}, Japan.\\
  $^3$ Graduate School of Information Science and Technology, The University of Tokyo, 7-3-1, Hongo, Bunkyo-ku,\\ Tokyo \emph{113-8656}, Japan.\\
  $^4$ Center for Advanced Intelligence Project, RIKEN, 1-4-1, Nihonbashi, Chuo-ku, Tokyo \emph{103-0027}, Japan.\\
  $^5$ Interfaculty Initiative in Information Studies, The University of Tokyo, 7-3-1, Hongo, Bunkyo-ku, \\ Tokyo \emph{113-0033}, Japan.\\
  $^6$ National Research Institute for Earth Science and Disaster Resilience, 3-1, Tennodai, Tsukuba-shi, \\Ibaraki \emph{305-0006}, Japan.\\
  }
\date{Received 1998 December 18; in original form 1998 November 22}
\pagerange{\pageref{firstpage}--\pageref{lastpage}}
\volume{200}
\pubyear{1998}

\begin{document}

\label{firstpage}

\maketitle

\begin{summary}
Uncovering the distribution of magnitudes and arrival times of aftershocks is a key to comprehending the characteristics of earthquake sequences, which enables us to predict seismic activities and conduct hazard assessments. However, identifying the number of aftershocks immediately after the main shock is practically difficult due to contaminations of arriving seismic waves. To overcome this difficulty, we construct a likelihood based on the detected data, incorporating a detection function to which Gaussian process regression (GPR) is applied. The GPR is capable of estimating not only the parameters of the distribution of aftershocks together with the detection function, but also credible intervals for both the parameters and the detection function. The property that the distributions of both the Gaussian process and aftershocks are exponential functions leads to an efficient Bayesian computational algorithm to estimate hyperparameters. After its validation through numerical tests, the proposed method is retrospectively applied to the catalog data related to the 2004 Chuetsu earthquake for the early forecasting of the aftershocks. The results show that the proposed method stably and simultaneously estimates distribution parameters and credible intervals, even within $t\leq 3$h after the main shock. 
\end{summary}


\section{Introduction}

A massive earthquake triggers a number of aftershocks. The classical representative models to describe the temporal distribution of aftershocks were established as the Omori-Utsu \citep{Omori1894, Utsu1961} and the Gutenberg-Richter \citep{Gutenberg1944} formulae, the latter of which also considers information on magnitudes. \citet{Ogata1988_2} proposed, having extended the Omori-Utsu formula, the Epidemic Type Aftershock Sequence (ETAS) model to describe more realistically  that large aftershocks also excite subsequent aftershocks, similar to the main shock. The distribution of aftershocks enables forecasting  seismic activities and conducting hazard assessments \citep{Resenberg1989, Resenberg1994, Kagan2000}. Until now, previous studies have proposed many statistical methods to estimate the parameters involved in the models \citep{Aki1965, Ogata1983, Ogata1988_2}. A disadvantage of these parameter estimation methods is that they assume the existence of a complete dataset without missing values. However, detecting all aftershocks immediately after the main shock is unrealistic due to contaminations by a tremendous amount of seismic waves. Such incomplete data cause underestimations in the counting of aftershocks at the time-dependent completeness magnitude.

In statistics, situations in which only some detected data are available are known as biased sampling problems \citep{Vardi1982, Vardi1985}. In our case, the detection probability of aftershocks clearly depends on the magnitudes and elapsed time from the main shock. This type of biased sampling data is termed as “missing not at random” (MNAR), in which the detection probability depends on the values of undetected data. Introducing a detection function, which is a model of the detection probability, enables correcting the bias \citep{Qin2017}.  Several studies have tackled this problem by estimating the time-dependent completeness magnitude \citep{Hainzl2016} and introducing a parametric model: the cumulative distribution function (CDF) of a normal distribution \citep{Ringdal1975, Ogata1993, Ogata2006, Omi2013, Omi2014, Omi2015_2, Omi2015, Zhuang2017, Martinsson2018} and CDF of an exponential distribution with an upper limit \citep{Mignan2012, Kijko2017, Mignan2019}. The detection function enables the construction of valid estimators in MNAR. Three problems remain to be solved: (i) the resulting estimators are often unstable; (ii) misspecification of the detection function causes bias; (iii) estimation of the detection function is difficult even with a correct model. Problem (i) arises from the simultaneous estimation of the detection function and the distribution of aftershocks. Problem (ii) results from the fact that the bias correction strongly depends on how close the defined detection function is to the true one. Problem (iii) is because some integration is required in the likelihood, which makes estimations difficult in biased sampling problems. In this study, we propose a nonparametric Bayesian estimator to overcome these three problems. Appropriate prior information, considering characteristics of seismic activities in a target area, solves problem (i). A modeling of the detection function based on the technique of the Gaussian Process Regression (GPR),  which enables us to estimate an arbitrary continuous function from a given dataset without assuming a specific functional form, solves problems (i) and (ii) simultaneously. The GPR has been accepted widely in recent machine learning research due to its flexibility and wide coverage of function spaces \citep{Rasmussen2006, Matthews2018}. As for the computation of the model parameters, we propose an efficient Bayesian estimation algorithm utilizing the fact that the distributions of both the Gaussian process (GP) and aftershocks are exponential functions, which are compatible with computation, solving the third problem (iii). 

Another advantage of the GPR is that it is capable of evaluating uncertainties or credible intervals of the estimated parameters naturally, which has been difficult in previous studies, in spite of knowing that uncertainty is inevitable in making statistical decisions. In summary, the proposed method can solve the three problems mentioned above and additionally estimate the uncertainties of the parameters.

The remainder of this paper is organized as follows. Section~\ref{sec:2} introduces the GPR and proposes a method to estimate parameters for the distribution of aftershocks with a detection function through the GPR. Section~\ref{sec:3} validates the proposed method through numerical tests. Section~\ref{sec:4} demonstrates the effectiveness of the proposed method by applying it to the catalog data of the 2004 Chuetsu earthquake. Section~\ref{sec:5} concludes the present study, including future perspectives.

\section{Methodology}
\label{sec:2}

The present paper proposes the use of a detection function based on the GPR to model temporal changes in the detection probability of aftershocks, even immediately after a main shock. This section first gives a brief explanation of the GPR and then introduces the proposed method, especially its theoretical properties and an efficient computational algorithm.

\subsection{Gaussian process regression}
Recent studies in the solid Earth sciences used the GPR to construct models from given data in the cases that the physical or chemical process that produced the data was unknown or too complex. The GPR estimates a regression function simultaneously with its uncertainty through Bayesian nonparametric estimation. For example, \citet{Kuwatani2018} proposed to apply the GPR to interpolate the observed quantities of chemical compositions along with the radius of a rock. The estimated uncertainty often provides valuable information for observational or experimental designs, such as a suggestion of times and/or places of the next new observations or measurements. The present study adopts the GPR to obtain a nonparametric regression function in the framework of Bayesian estimation.

\begin{figure}
  \centering
  \includegraphics[width=140mm]{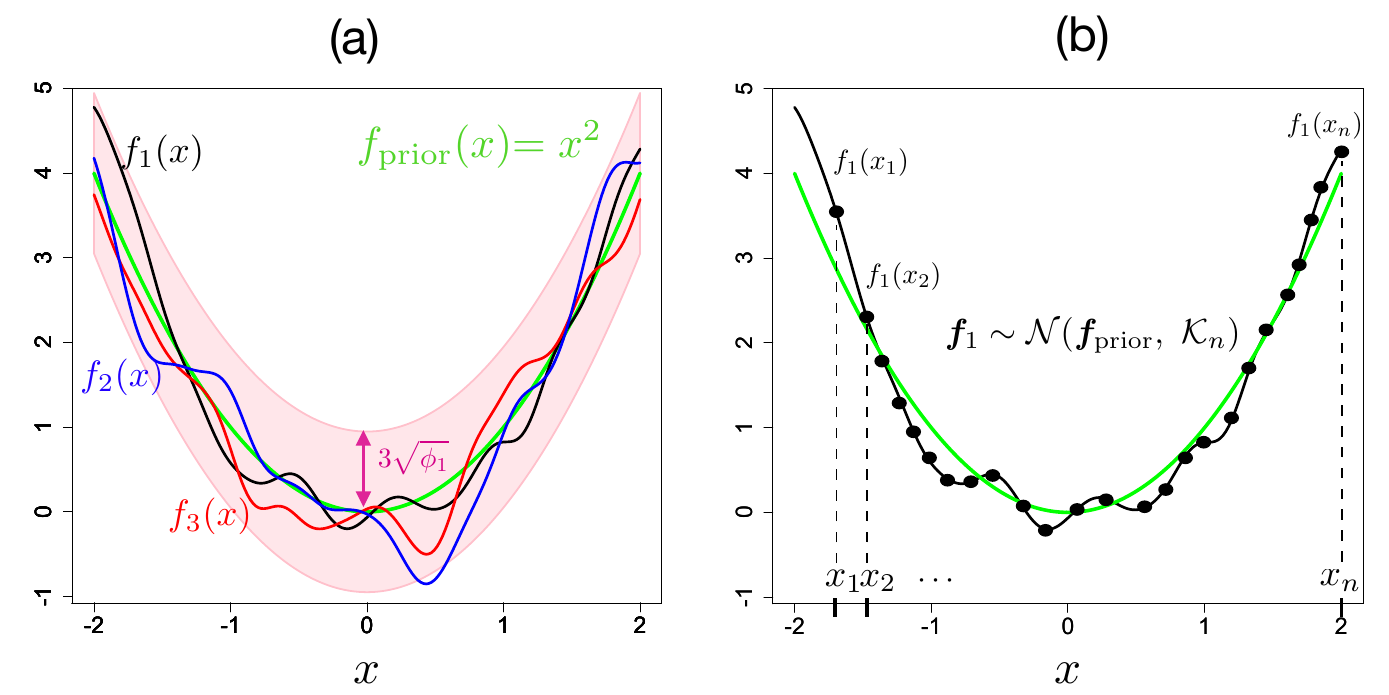}
  \caption{Example of GP: (a) three sampled functions $f_1(x)$ (black line), $f_2(x)$ (blue line), and $f_3(x)$ (red line) generated from a GP with $f_{\prior}(x)=x^2$ (green line), $\phi_1=0.3$, and $\phi_2=0.1$, and the red shaded zone is three times the standard deviation $(=3\sqrt{\phi_1})$ from the mean function; (b) values of the function $f_1(x)$ at arbitrary $n$ points $x_1, x_2, \ldots, x_n$. A set of these values forms a function sampled from the distribution of the GP when $n$ goes to infinity.}
\label{fig:gp}
\end{figure}

We first briefly explain the GP, on which the GPR is based. Fundamental statistics mention that, for a random variable, there exists a corresponding distribution that generates random ``values". The GP is an extension of this concept to a random function, i.e., there exists a corresponding distribution that generates random ``functions". The distribution of the GP is usually denoted as GP$(f_{\prior}(\cdot)$, $\mK(\cdot, \cdot) )$, where $f_{\prior}(\cdot)$ is the mean function and $\mK(\cdot, \cdot)$ is the variance function or ``kernel". The notation $f(\cdot)$ is used in this paper to abbreviate arguments, emphasizing that $f$ is a function. A radial basis function is often chosen among various candidate functions for the variance function or kernel $\mK(\cdot, \cdot)$:
 \begin{eqnarray}
 \mK(x_1, x_2)=\phi_1 \exp\llp-\frac{(x_1-x_2)^2}{\phi_2^2}\rrp\quad (\phi_1, \phi_2>0), \label{kernel}
 \end{eqnarray}
where $x_1$ and $x_2$ are arbitrary real numbers. Another reason why this study adopts the radial basis function for the kernel is that it adequately covers an infinite-dimensional function space with only a few hyperparameters. See \citet{Rasmussen2006} for a detailed explanation of the GP, including other kernel functions. Figure~\ref{fig:gp}(a) illustrates three random ``functions" generated from a GP$(f_{\prior}(\cdot)$, $\mK(\cdot, \cdot) )$ with $f_{\prior}(x)=x^2$, $\phi_1=0.3$ and $\phi_2=0.3$. For any $n$ points $x_1, x_2, \ldots, x_n$, where $n$ is an arbitrary positive integer, the GP is mathematically equivalent to $\bm{f}=(f(x_1), f(x_2), \ldots, f(x_n))^\top$ computed from a sampled function $f(\cdot)$ that follows a multivariate normal distribution with mean $\bm{f}_{\prior}=(f_{\prior}(x_1), f_{\prior}(x_2), \ldots, f_{\prior}(x_n))^\top$ and variance $\mK_n=(\mK(x_i, x_j))$. Here $(a_{ij})$ means a matrix having $a_{ij}$ in its $(i, j)$-th element, and the superscript $\top$ means transposition. Note that the variance of the value $f(x_i)$ at any point $x_i$ is $\mK(x_i, x_i)=\phi_1$, and triple of its standard error $3\sqrt{\phi_1}$ produces an approximately 99.7\% confidence interval for the GP shaded region in red in Figure~\ref{fig:gp}(a). 
Figure~\ref{fig:gp}(b) indicates that the values of the function $f_1(\cdot)$ (black line in Figure~\ref{fig:gp}(a)) at arbitrary $n$ points $x_1, x_2, \ldots, x_n$ follows a multivariate normal distribution $\mN( \bm{f}_1, \mK_{n})$, where $\bm{f}_1=(f_1(x_1), f_1(x_2), \ldots, f_1(x_n))^\top$. Intuitively speaking, when the number of points $n$ goes to infinity, a set of points forms the function $f_1(\cdot)$ that is a sample from the distribution of the GP. This consideration indicates that the GP is a stochastic process obtained by letting the dimension of a multivariate normal distribution go to infinity. 

\begin{figure}
  \centering
  \includegraphics[width=130mm]{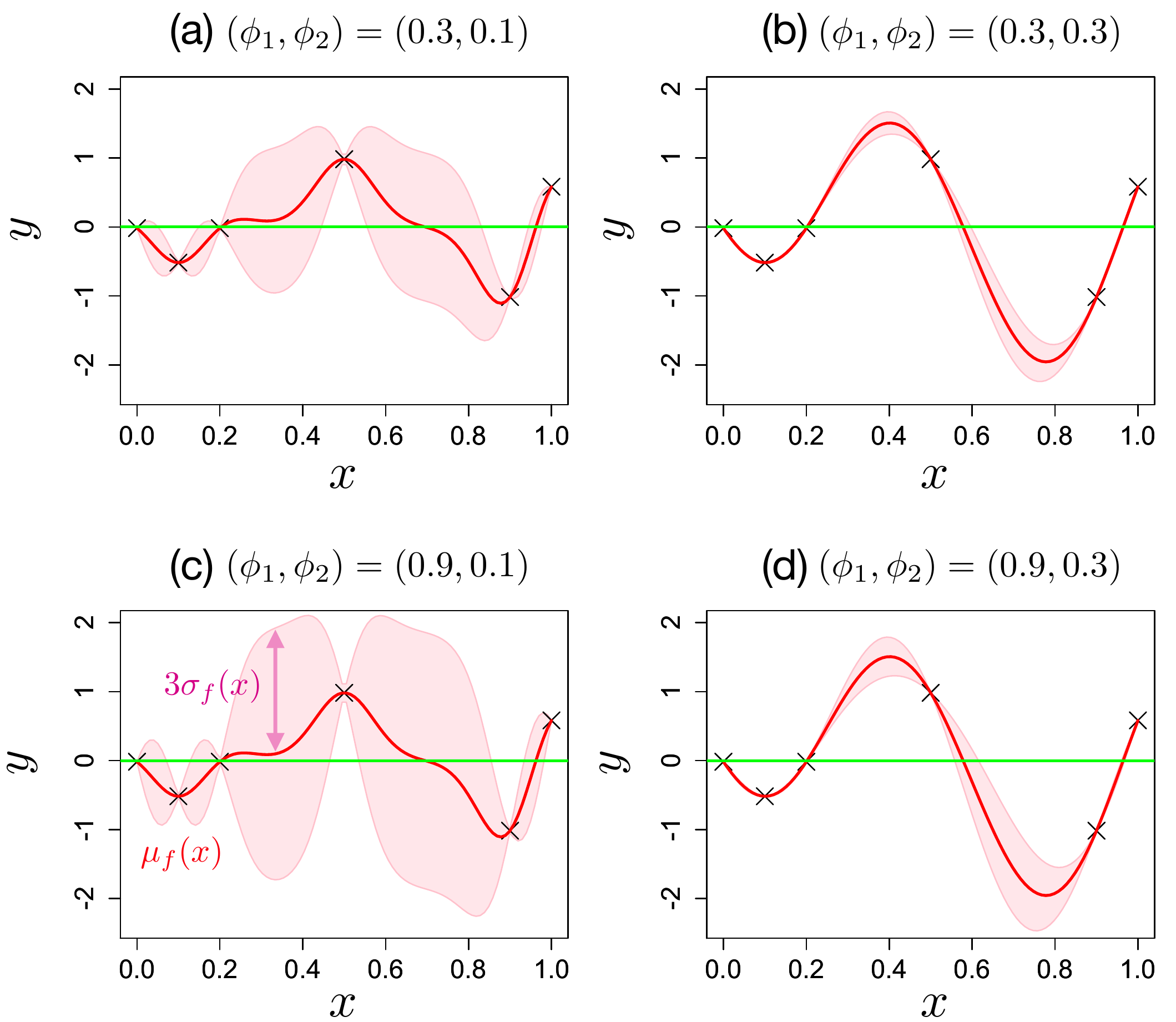}
  \caption{Gaussian process regression applied to six data points (``$\times$") with different hyperparameters $(\phi_1, \phi_2)$. The green line is the mean function $f_{\prior}(\cdot)$ of the prior distribution, which is assumed to be identically zero, the red curve is the estimated mean $\mu_f(\cdot)$ of the predictive distribution, and the shaded region is the deviation $3\sigma_f(\cdot)$ from the mean of the predictive distribution.}
  \label{fig1}
\end{figure}

The GPR is a Bayesian estimation method for a target function using a distribution of the GP as the prior information. Let $\bm{x}=(x_1, \ldots, x_n)^\top$ and $\bm{y}=(y_1, \ldots, y_n)^\top$ be sets of explanatory and response variables, respectively, and both are assumed to relate to each other through an unknown regression function $f(\cdot)$, i.e., $y_i = f(x_i)\;(i = 1, \ldots, n)$, where $n$ denotes the number of time points. The GPR estimates the function $f(\cdot)$ from a given dataset assuming the kernel function mentioned above. We assume GP$(f_{\prior}(\cdot)$, $\mK(\cdot, \cdot) )$ to be a prior distribution of the target function $f(\cdot)$. The mean function $f_{\prior}(\cdot)$ is often assumed to be identically zero  since the mean is adjustable by subtracting the sample mean of $\bm{y}$. Estimation of a function $f(\cdot)$ is equivalent to that of the value of $f(x^*)$ at any fixed point $x^*$, where the superscript ``$\ast$" is used to discriminate the fixed point from the data points. A posterior distribution of $f^*=f(x^*)$ given a dataset $\mathcal{D}=\{x_1, \ldots, x_n, y_1, \ldots, y_n\}$ is called the predictive distribution. The law of total probability yields the predictive distribution as
\begin{eqnarray}
p(f^*\mid x^*, \mathcal{D})= \int p(f^*\mid x^*, \bm{f}, \mathcal{D})p(\bm{f}\mid \mathcal{D})d\bm{f}, \label{predictive}
\end{eqnarray}
where $p(f^*\mid x^*, \bm{f}, \mathcal{D})$ is the conditional probability density function (PDF) of $f^*$ with the given fixed point $x^*$, unobserved $\bm{f}$, and the dataset $\mD$, and the PDF $p(\bm{f}\mid \mD)$ is the prior on $\bm{f}$ given by a multivariate normal distribution $\mN( \bm{f}, \mK_{n})$ as mentioned above. The predictive distribution becomes a normal distribution again in accordance with the reproductive property. Therefore, the predictive distribution for a point $x^*$ is a normal distribution with mean $\mu_f(x^*)$ and variance $\sigma^2_f(x^*)$ computed as
\begin{eqnarray}
\mu_f(x^*)=\bm{\kappa}^\top_* \mK^{-1}_n\bm{y}, \quad  \sigma^2_f(x^*)= \kappa_{**}-\bm{\kappa}^\top_*\mK^{-1}_n\bm{\kappa}_*,
\end{eqnarray}
where $\mK_n=(\mK(x_i, x_j))$, $\bm{\kappa}_*=\{\mK(x^*, x_1), \ldots, \mK(x^*, x_n)\}^\top$ and $\kappa_{**}=\mK(x^*, x^*)$. The maximization of the marginal likelihood
\begin{eqnarray}
\prod_{i=1}^n p(y_i\mid x_i;\phi)=\prod_{i=1}^n\int p(y_i\mid x_i, \bm{f}; \phi)p(\bm{f}\mid x_i;\phi)d\bm{f} \label{marginal}
\end{eqnarray}
determines the hyperparameters $\phi= (\phi_1, \phi_2)^\top$, where the integration on the right-hand side is explicitly computable since the integrand given as the product of normal distributions $p(y_i\mid x_i, \bm{f}; \phi)$ and $p(\bm{f}\mid x_i;\phi)$ is again a normal distribution. Figure~\ref{fig1} shows the predictive distributions estimated from six data points with changing hyperparameters $\phi_1$ and $\phi_2$ in the kernel (eq.~\ref{kernel}). As mentioned above, the GPR successfully obtains not only the mean function but also its standard error, i.e., credible interval. Figure~\ref{fig1} also indicates that $\phi_1$ and $\phi_2$ strongly associate with scale and shape of the regression function, respectively. A comparison between Figures~\ref{fig1}(a) and \ref{fig1}(c), or \ref{fig1}(b) and \ref{fig1}(d)indicates that small/large $\phi_1$ means a small large/credible interval. Another comparison between Figure~\ref{fig1}(a) and \ref{fig1}(b), or \ref{fig1}(c) and \ref{fig1}(d) shows that small/large $\phi_2$ means an oscillating/smoothed regression function.
These results indicate the importance of deciding the hyperparameters.

\subsection{Notation and models}
 According to the Omori-Utsu law, the aftershock occurrence rate $n(t)$ at elapsed time $t$ from the main shock follows a non-stationary Poisson process \citep{Omori1894, Utsu1961}:
\begin{eqnarray}
n(t; \bm{\tau})=\frac{K}{(t+c)^p},  \label{Omori}
\end{eqnarray}
where $\bm{\tau}$ is a vector containing all the model parameters, i.e., $\bm{\tau}=(K, p, c)^\top$. The parameter $K$ controls the level of seismic activity, i.e., large/small $K$ reflects a large/small number of aftershocks. The parameter $p$ is the slope of the occurrence rate on the logarithmic scale. 
The parameter $c$ characterizes the length of “capped time”, which indicates the well-known phenomenon that the occurrence rate is below some level for a while immediately after the main shock \citep{Utsu1961, Ogata1983}. According to the Gutenberg-Richter law, the intensity rate of the magnitude $M$ is described by an exponential function \citep{Gutenberg1944}:
\begin{eqnarray}
m(M; b)=A10^{-b M}\propto \exp(-\beta M), \label{M_domain}
\end{eqnarray}
where $A$ and $b$ (or $\beta=b\ln 10$) are constants. The parameter $b$ is of most interest since it reflects the intensity rate of the magnitude in the statistical meaning. Combining eqs.~\eqref{Omori} and \eqref{M_domain}, the joint occurrence rate of aftershocks as a function of elapsed time $t$ and magnitude $M$ is represented by the product of $n(t)$ and $m(M)$ as
\begin{eqnarray}
\lambda(t, M; \bm{\tau}, \beta)=\frac{K'}{(t+c)^p}\beta e^{-\beta (M-M_0)}, \label{joint}
\end{eqnarray}
where $K'=\beta^{-1}KA\exp(-\beta M_0)$ and $M_0$ is the magnitude of the main shock used to adjust the scale of $K'$ \citep{Utsu1970}. Note that a unique decomposition into $K$ and $A$ is mathematically impossible, even if $K'$ is obtained. \citet{Resenberg1989, Resenberg1994} pointed out that an estimation of $K'$ leads to a forecast of seismic activities, so that the unique decomposition problem does not matter in this context. Hereafter, we use $K$ instead of $K'$ for notational simplicity.

\begin{figure}
  \centering
  \includegraphics[width=150mm]{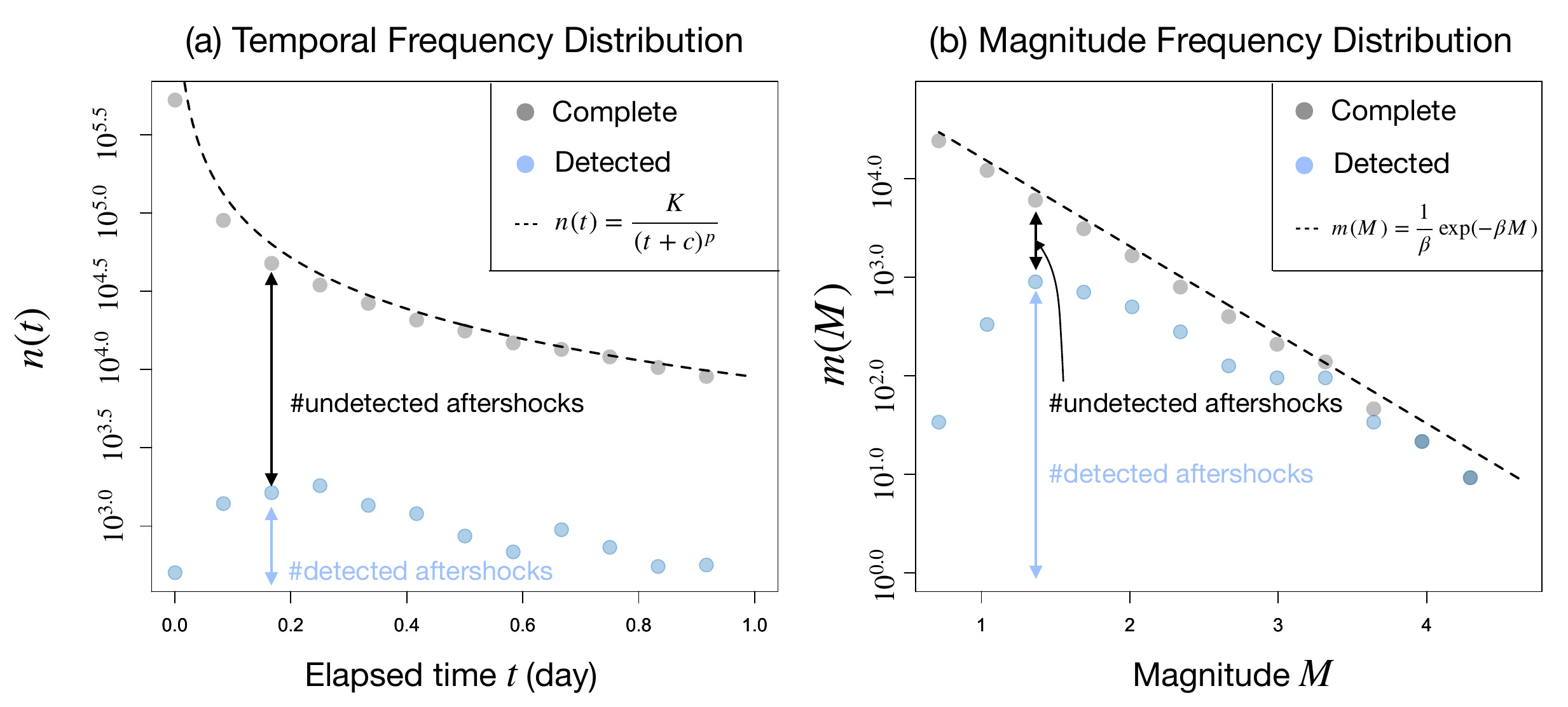}
  \caption{Frequency distribution of (a) elapsed times, and (b) magnitudes on complete and detected aftershocks. Complete data of aftershocks are artificially generated by the Omori-Utsu and the Gutenberg-Richter laws,  
and detected aftershocks are generated by removing some of the complete data with the probability law. }
  \label{new_fig1}
\end{figure}

An exact count of all aftershocks right after a main shock is very complicated due to contaminations by a number of arriving seismic waves, so that the occurrence rate of aftershocks is almost always underestimated. Figure \ref{new_fig1} shows a comparison of artificially generated aftershocks with detected aftershocks, and illustrates how undetected aftershocks distort both the occurrence rate (left) and the distribution of magnitudes (right).  

 In order to correct the bias, the present study adopts the probit-type detection function, as used in \citet{Ringdal1975}, \citet{Ogata1993}, \citet{Ogata2006}, \citet{Omi2013,Omi2014,Omi2015_2,Omi2015}, and \citet{Martinsson2018}:
 \begin{eqnarray}
 \pi(t, M;\mu, s) &=& P(\delta=1\mid t, M; \mu, s)\nonumber\\
 &=& \int_{-\infty}^M\frac{1}{\sqrt{2\pi s^2}}\exp\llp -\frac{(x-\mu(t))^2}{2s^2}\rrp dx, \label{detection}
 \end{eqnarray}
 where $\delta$ is a detection indicator that takes 1/0 if the aftershock is detected/undetected, $\mu(t)$ is the magnitude that makes aftershocks detectable with a probability of 50\% at elapsed time $t$, and $s$ is a scale parameter that determines the steepness of the exponential curve. The detection function $\pi(t, M)$ is the cumulative distribution of a Gaussian
normal distribution, and $s$ is the standard deviation of this distribution. Roughly speaking, the function $\mu(t)$ is decreasing with respect to the elapsed time $t$ because only large aftershocks are detectable immediately after the main shock and even small aftershocks are detectable after enough time has passed.

To distinguish the notations related to complete and detected data, let $t_1$ and $M_1$ be the elapsed time and magnitude of a detected aftershock, respectively. The subscript ``1" indicates the detected data, i.e., $\delta=1$. Note that $t_1$ and $M_1$ are always available, although the complete data $t$ and $M$ may be unavailable.
Supposing that $n$ aftershocks are detected, let the pair of elapsed time from the main shock and the magnitude of the $i$-th aftershock be $(t_{1i}, M_{1i})\;(i=1, \ldots, n)$. The thinning operation or the random deletion in point processes \citep{Ogata1993} yields the likelihood function for the detected magnitudes
 \begin{eqnarray}
&&\prod_{i=1}^nL(M_{1i}\mid t_{1i}; \beta, \mu, s^2)\nonumber\\
&&=\prod_{i=1}^n\frac{e^{-\beta M_{1i}}\pi(t_{1i}, M_{1i}; \mu, s^2)}{\int_{-\infty}^\infty e^{-\beta M}\pi(t_{1i}, M; \mu, s^2)dM}\nonumber \\
&&=\prod_{i=1}^n\beta \exp\llp -\beta(M_{1i}-\mu(t_{1i}))-\frac{1}{2}\beta^2 s^2\rrp \pi(t_{1i}, M_{1i}; \mu, s^2), \label{detect_data}
 \end{eqnarray}
 where the function $\mu(\cdot)$ is assumed to be known, although it is estimated later in practice. Using the estimated $\hat{\beta}$ and $\hat{s}$ obtained by maximizing eq.~\eqref{detect_data}, $\bm{\tau}$ can be estimated by maximizing the log-likelihood function for detected elapsed times within any time interval $(0, ~T)$ \citep{Ogata1993},
\begin{eqnarray}
\ln L=\sum_{0<t_{1i}<T}\ln \nu(t_{1i}; \bm{\tau}, \mu(t_{1i}), \hat{\beta}, \hat{s})-\int_0^T \nu(t; \bm{\tau}, \mu(t), \hat{\beta}, \hat{s})dt, \label{loglike_tau}
\end{eqnarray}
where $\nu(t)$ is an intensity function for detected data defined by
\begin{eqnarray}
&&\nu(t; \bm{\tau}, \mu(t), \beta, s) \nonumber\\
&& =\int_{-\infty}^{\infty}\lambda(t,M; \bm{\tau}, \beta)\pi(t,M; \mu, s)dM\nonumber\\
&& =\frac{K}{(t+c)^p}\exp\llp-\beta(\mu(t)-M_0)+\frac{1}{2}\beta^2s^2\rrp. \label{nu2}
\end{eqnarray}

However, the serious problem that $\mu(\cdot)$ is unknown remains. In frequentist ways, \citet{Martinsson2018} and \citet{Mignan2019} proposed a mixture detection function of parametric models. Although the mixture detection function is flexible and  approximates the true detection function well, either misspecification of the parametric models or the number of mixture components causes biased estimates. As for nonparametric models, \citet{Ogata1993} applied a B-spline basis function to $\mu(\cdot)$, and \citet{Ogata2006} proposed a specific parametric model based on the 2003 Miyagi-Ken-Oki earthquake as
\begin{eqnarray}
\mu(t)=a_0+a_1 \exp\llp-\alpha (\lceil -\ln t_{11} \rceil+\ln t)^\gamma\rrp,  \label{Ogata}
\end{eqnarray}
where $a_0$, $a_1$, $\alpha$, and $\gamma$ are parameters to be estimated and $\lceil \cdot \rceil$ is the ceiling function. Recently, \citet{Omi2013} proposed a flexible nonparametric Bayesian estimation. They assumed a prior on $\mu(t_{1i})\;(i=1, \ldots, n)$ as 
\begin{eqnarray}
(\mu_1, \ldots, \mu_n) \sim p(\mu_1, \mu_2) \prod_{i=1}^{n-2}\frac{1}{\sqrt{2\pi V}}\exp\lp-\frac{(\mu_{i+2}-2\mu_{i+1}-\mu_i)^2}{2V}\rp,
\end{eqnarray}
where $V$ is a hyperparameter and $p(\mu_1, \mu_2)$ is a constant, and $\mu_i (i=1, \ldots, n)$ is estimated by the posterior mean. This prior indicates that the mean of each  $\mu_n$ is a point apart from $\mu_{n-1}$ with a distance of $(\mu_{n-1}-\mu_{n-2})$ given the values $\mu_{n-1}$ and $\mu_{n-2}$. Their method does not require any specification of $\mu(\cdot)$ and it can naturally incorporate the prior information of $\beta$, which makes estimates considerably stable. The posterior distribution of $\beta$ can be estimated in the same aftershock area that occurred before the main shock. Also, it is difficult to evaluate the credibility of $\mu(\cdot)$.

\subsection{Estimation of $\beta$, $s^2$, and $\pi(\cdot)$}
\label{2.2}

\begin{figure}
  \centering
  \includegraphics[width=120mm]{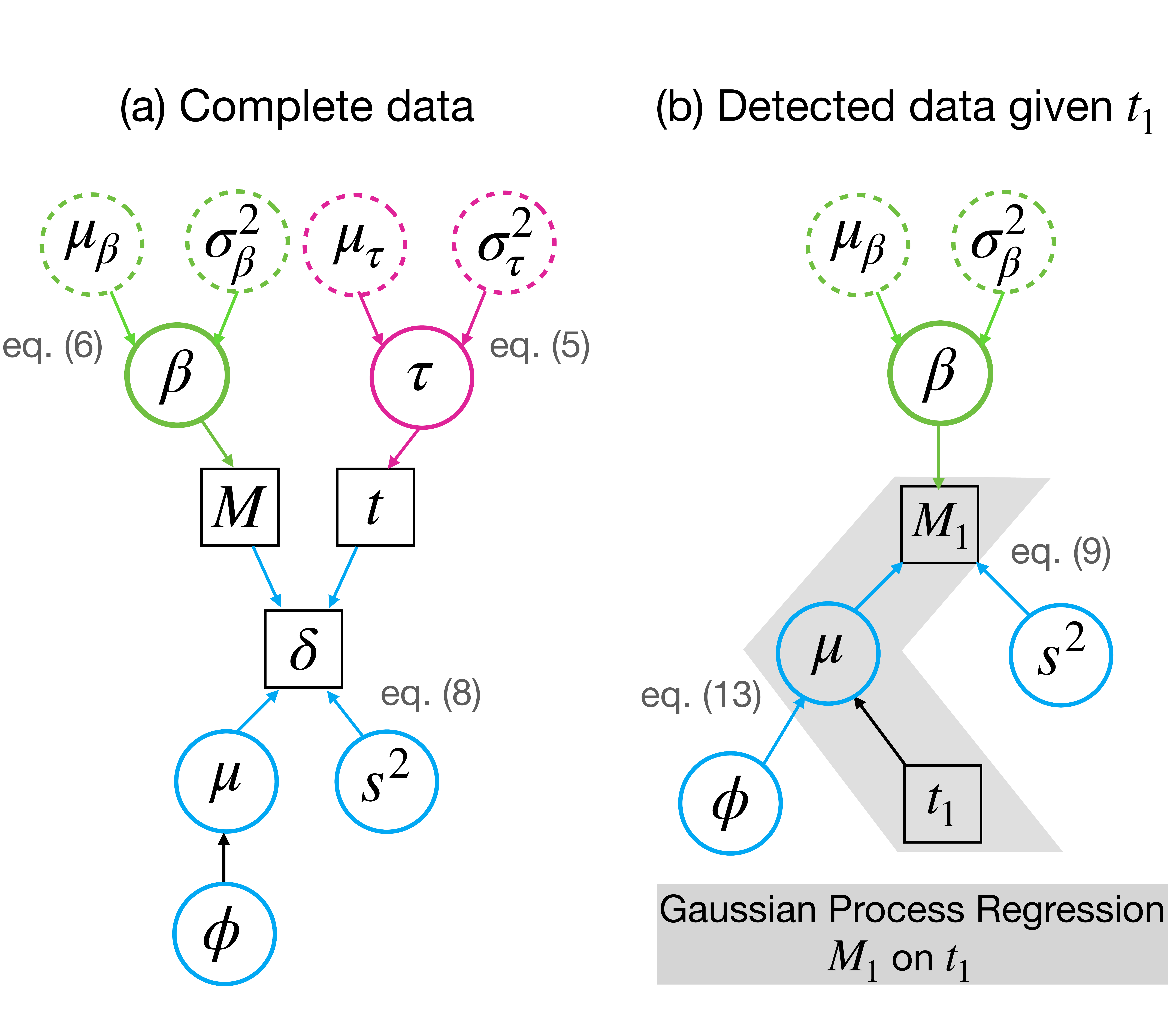}
  \caption{Graphical models on (a) complete data; (b): detected data given $t_1$. Each square, circle, and dotted circle indicates data, parameters, and subjective hyperparameters, respectively. Each green, red, and blue color indicates the relation by the Gutenberg-Richter law, the Omori-Utsu law, and the detection function, respectively.}
\label{fig_structure}
\end{figure}

Figure~\ref{fig_structure}(a) shows a graphical model for complete data, which illustrates how the data, parameters, and hyperparameters relate to each other in accordance with the Omori-Utsu law, the Gutenberg-Richter law, and the detection function. 
Since the complete data are not available, we construct a likelihood based on the detected data $M_1$ given $t_1$, as shown in Figure~\ref{fig_structure}(b). The distribution of $M_1$ given $t_1$ is already derived in eq.~\eqref{detect_data}. Once the graphical model for the detected data is obtained, it can be realized that the relation among $M_1$, $\mu$, and $t_1$ is exactly the same as in the regression: each input $x$, the regression $f(\cdot)$, and output $y$ corresponds to $t_1$, $\mu(\cdot)$, and $M_1$, respectively. Based on this idea, we put the GP prior on $\mu(\cdot)$ and consider a nonparametric Bayesian estimation. However, unlike the GPR explained in Section 2.1, the distribution derived in eq.~\eqref{detect_data} is not a normal distribution, and the predictive distribution shall be more complicated.

In this study, we put a GP prior on the ``function" $\mu(\cdot)$, not on the ``points" $\mu(t_{1i})\;(i=1, \ldots, n)$. The GP prior leads to an explicit form of the posterior distribution of the hyperparameters and the predictive distribution of $\mu(\cdot)$, as seen in the following discussion. The prior for $\mu(\cdot)$ can be denoted as
\begin{eqnarray}
\mu \sim \mathrm{GP} (\mu_{\prior}(\cdot), \mK(\cdot, \cdot)), \label{new_mu}
\end{eqnarray}
where $\mK$ is defined as 
\begin{eqnarray}
\mK(x_1, x_2)=\phi_0+\phi_1 \exp\llp-\frac{(x_1-x_2)^2}{\phi_2^2}\rrp\quad (\phi_1, \phi_2>0). \label{new_kernel}
\end{eqnarray}
The first term $\phi_0$ in eq.~\eqref{new_kernel} is added to the usual kernel eq.~\eqref{kernel} to make the matrix $\mK^{-1}$ always nonsingular. A subjectively small value, e.g., $10^{-7}$, is set to $\phi_0$ for a stable estimation of the other parameters.
The prior $\mu_{\prior}(\cdot)$ should be a function except for identically zero in this case since adjusting the mean of the prior as in the standard GPR is impossible. We propose to use the Ogata model (eq. \ref{Ogata}) because of the prior information of  $\mu(\cdot)$ since the model successfully represents the detection function on average, although representing small oscillations is difficult due to the limited number of parameters \citep{Omi2013}.

In summary, the parameters to be estimated are $\bm{\theta}=(\beta, s^2, \phi_1, \phi_2)^\top$. 
We apply exactly the same subjective priors as \citet{Omi2015} that are estimated in \citet{Omi2015_2} by fitting the Omori-Utsu and the Gutenberg-Richter laws to 38 aftershock sequences in Japan from the JMA catalog. Let $\beta\sim p(\beta)=\mN(1.96,~0.34^2)$ be the estimated prior distribution for $\beta$. The posterior distribution of $\bm{\theta}=(\beta, s^2, \phi_1, \phi_2)^\top$ and the predictive distribution of $\mu(\cdot)$ are computable as follows (see Appendix for the proofs).
\begin{itemize}
	\item Posterior distribution of $\bm{\theta}=(\beta, s^2, \phi_1, \phi_2)^\top$:
	\begin{eqnarray}
	&&p(\bm{\theta}\mid \mD)\nonumber\\
&&\propto p(\beta)\int p(\bm{M}_1\mid \bm{t}_1; \beta, \bm{\mu}, s^2)p(\bm{\mu}\mid \bm{t}_1; \phi_1,\phi_2)d\bm{\mu} \nonumber\\
&&
\begin{split}
&=\beta^n \exp\llp -\beta (\bm{M}_{1}-\bm{\mu}_{\prior})^\top\bm{1}_n-\frac{1}{2}\beta^2 \bm{1}_n^\top\lp s^2 I_n-\mK_{n}\rp \bm{1}_n\rrp\\
&\quad \times  p(\beta) \int_{\mM}  \mathcal{N}\lp\bm{x};~ \tilde{\bm{\mu}}_n, \tilde{\mK}_n\rp d\bm{x}, \end{split}
\label{marginal}
	\end{eqnarray}	
	where $\mM=\otimes_{i=1}^n\{x_i\leq M_{1i}\}$ is the integration interval, $\otimes_{i=1}^n$ is the direct product of the $n$ sets $\{x_i\leq M_{1i}\}\;(i=1, \ldots, n)$, $\tilde{\bm{\mu}}_n=\bm{\mu}_{\prior}+\beta \mK_n\bm{1}_n$, $\bm{1}_n$ is the $n$-dimensional vector that has one in all of its elements, $\tilde{\mK}_n=\mK_n+s^2 I_n$,  $I_n$ is the $n$-dimensional identity matrix, and $\mN(\bm{x}; \bm{\mu}, \Sigma)$ is the PDF of the normal distribution, with mean $\bm{\mu}$ and variance $\Sigma$. One reasonable estimator of $\bm{\theta}$ is the unique maximizer of eq. \eqref{marginal}, called a maximum a posteriori (MAP) estimator.
	
	\item Predictive distribution of $\mu(\cdot)$:
	\begin{eqnarray}
&&p(\mu^*\mid t^*_1, \mathcal{D})\nonumber\\
 &&=  \int p(\mu^* \mid  t_1^*, \bm{\mu}, \mD)p(\bm{\mu}\mid \mD) d\bm{\mu} \label{pred0}\\
&&=\bE_\trunc\llp\mathcal{N}\lp \mu^*;~D^*(\bm{X}),~(\tau^*)^2\rp\rrp, \label{pred}
	\end{eqnarray}
	where 
	\begin{eqnarray}
	D^*(\bm{x})=(\bm{x}+s^2 \mK_n^{-1}\tilde{\bm{\mu}}_n)^\top  \llp (\tau^*)^2 \tilde{\mK}_n^{-1}+\tilde{\mK}_n^{-1}\bm{\kappa}_*\bm{\kappa}_*^\top\tilde{\mK}_n^{-1} \rrp\frac{\bm{\kappa}_*}{\kappa_{**}},\label{var_eq}
	\end{eqnarray}
	$\bm{\kappa}_*=(\mK(t_1^*, t_{11}), \ldots, \mK(t_1^*, t_{1n}))^\top$, $\kappa_{**}=\mK(t_1^*, t_1^*)$, and $(\tau^*)^2=\kappa_{**}-\bm{\kappa}_*^\top \tilde{\mK}^{-1}_n\bm{\kappa}_*$. The expectation $\bE_\trunc$ is on a variable $\bm{X}\sim\mathcal{TN}(\tilde{\bm{\mu}}_n, \tilde{\mK}_n, \mathcal{M})$, where $\mathcal{TN}(\bm{\mu}, \Sigma, \mathcal{R})$ is the truncated multivariate normal distribution with mean $\tilde{\bm{\mu}}_n$, variance $\tilde{\mK}_n$, bounded by the region $\mathcal{R}$, 
In particular, the mean and the variance of the predictive distribution are $D^*(\bm{\xi})$ and $(\tau^*)^2$,
	where 
\begin{eqnarray}
	\bm{\xi}=\bE_\trunc(\bm{X}). \label{xi}
\end{eqnarray} 
The symbol ``$\ast$" explicitly represents that the variable depends on $t^*_1$.
	\item Predictive distribution of detection probability $\pi(\cdot)$:
	\begin{eqnarray}
	&&\pi^*(M_1^*, t_1^*)\nonumber\\
	&&= \int P(\delta=1\mid M^*_1, \mu^*, \mD)  p(\mu^*\mid t^*_1, \mD)d\mu^* \nonumber\\
	&&=\bE_\trunc\llp \Psi \lp\frac{M^*_1-D^*(\bm{X})}{\sqrt{s^2+(\tau^*)^2}}\rp\rrp, \label{pred_detect}
	\end{eqnarray}
	where $\Psi(\cdot)$ is the cumulative distribution function of the standard normal distribution, i.e., $\mN(0, 1)$.
\end{itemize}

The integrations in eqs.~\eqref{pred},  \eqref{xi}, and \eqref{pred_detect} are computable by the Monte Carlo method since they involve expectations on the known truncated multivariate normal distribution $\mathcal{TN}(\tilde{\bm{\mu}}_n, \tilde{\mK}_n, \mathcal{M})$. However, computation of the integration in the posterior distribution (eq.~\ref{marginal}) is more difficult. We provide an efficient computational algorithm to obtain the posterior samples of the hyperparameter $\bm{\theta}$, instead of maximizing the posterior distribution (eq. \ref{marginal}) directly.

\subsection{Estimation of $\bm{\tau}$}

Recall that the log-likelihood function for $\bm{\tau}$ is given in eq.~\eqref{loglike_tau}, but $\mu(\cdot)$ is unknown in the intensity function (eq.~\ref{nu2}). \citet{Ogata1993} and \citet{Omi2013} replaced $\mu(\cdot)$ with the MAP estimator. We propose to use an estimated mean of the predictive distribution for $\mu(\cdot)$ (eq. \ref{xi}). For example, $(\mu(t_{1i}), \ldots, \mu(t_{1n}))$ is estimated by $(\hat{\xi}_1, \ldots, \hat{\xi}_n)$, where $\hat{\bm{\xi}}=\hat{\bE}_\trunc(\bm{X})$ is an estimated mean of the predictive distribution with its sample mean generated from $\mathcal{TN}(\tilde{\mu}_n,~\tilde{\mathcal{K}}_n, \mathcal{M})$. We also apply the following priors for $\bm{\tau}$ estimated in \citet{Omi2015_2} as explained in Section 2.3:
\begin{eqnarray}
\ln K\sim \mN(-4.86,~1.60^2),\quad p\sim \mN(1.05,~0.13^2),\quad \ln c\sim \mN(-4.02,~1.42^2). \label{prior2}
\end{eqnarray}
Let $p(\bm{\tau})$ be product of the three distributions in eq. \eqref{prior2}. Then, our estimating equation for $\bm{\tau}$ is 
\begin{eqnarray}
\log p(\bm{\tau})+\sum_{0<t_{1i}<T}\ln \nu(t_{1i}; \bm{\tau}, \hat{\xi}_i, \hat{\beta}, \hat{s})-\int_0^T \nu(t; \bm{\tau}, \mu(t), \hat{\beta}, \hat{s})dt. \label{loglike_tau}
\end{eqnarray}
As for the integration of the third term in eq.~\eqref{loglike_tau}, we use a left Riemann sum, discretizing the interval $[0,~T]$ into $10^4$ meshes having the same interval. The predictive distribution of $\mu(\cdot)$ enables us to estimate corresponding $(\mu_1, \ldots, \mu_{10^4})$, including data points except for detected aftershocks in a similar way to estimate $\bm{\xi}$, unlike the model by \citet{Omi2014}, in which only $(\mu(t_{1i}), \ldots, \mu(t_{1n}))$ is estimable. Our method can precisely evaluate the integration in the third term in eq.~\eqref{loglike_tau}.

\subsection{Computational algorithm}
Both the posterior distribution and the predictive distribution involve integration on the multivariate normal distribution over an $n$-dimensional hyperrectangle $\mM$. Since an explicit computation of the integral is difficult, even when the dimension is low \citep{Genz2009}, we divide the estimation into two steps: (i) estimation of hyperparameters $\bm{\theta}$; (ii) computation of mean and variance of the predictive distribution. 

In step (i), the data-augmentation method \citep{Tanner1987} is applied to obtain samples from the posterior distribution of $\bm{\theta}$. Regarding variable $\bm{x}$ in eq.~\eqref{marginal} as a latent variable yields another representation of the posterior distribution as
\begin{eqnarray}
p(\bm{\theta}\mid \mD) = \int_{\mM} p(\bm{\theta}, \bm{x}\mid \mD)d\bm{x},
\end{eqnarray}
where
\begin{eqnarray}
&&p(\bm{\theta}, \bm{x}\mid \mD)\nonumber\\
&&=\beta^n \exp\llp -\beta (\bm{M}_{1}-\bm{\mu}_{\prior})^\top\bm{1}_n-\frac{1}{2}\beta^2 \bm{1}_n^\top\lp s^2 I_n-\mK_{n}\rp \bm{1}_n\rrp  \mathcal{N}\lp\bm{x};~ \tilde{\bm{\mu}}_n, \tilde{\mK}_n\rp.
\end{eqnarray}
In accordance with the Gibbs sampling, augmented data can be sampled via the following two steps:
\begin{enumerate}
	\item Draw $\bm{x}\sim p(\bm{x}\mid \bm{\theta}, \mathcal{D})=\mathcal{TN}(\bm{x}; \tilde{\bm{\mu}}_n, \tilde{\mK}_n, \mathcal{M})$;
	\item Draw $\bm{\theta}\sim p(\bm{\theta}\mid \bm{x}, \mathcal{D})\propto p(\mathcal{D}, \bm{x}\mid \bm{\theta})p(\bm{\theta}\mid \bm{\theta}_0)$,
\end{enumerate}
 We apply the Gibbs sampling to obtain samples from the truncated normal distribution \citep{Geweke1991, Geweke2005} and use the package \textsf{tmvtnorm} of the R programming language \citep{tmvtnorm} in the practical programming. 
As for sampling of $\bm{\theta}$, the classical random walk MCMC (Markov Chain Monte Carlo) method \citep{Metropolis1953} works with a normal distribution as the proposal distribution. By repeating the two steps alternately, the obtained $\bm{\theta}$ becomes samples from the posterior of $\bm{\theta}$. Hyperparameters are estimated as the median of obtained samples. Once samples of $\bm{x}$ are obtained, $\bm{\xi}=\bE_\trunc(\bm{X})$ can be estimated by its sample mean. The mean of the predictive distribution of $\mu(\cdot)$ is computable with the estimated $\bm{\xi}$ by $D^*(\bm{\xi})$ given in eq.~\eqref{var_eq} and the variance is computable by $(\tau^*)^2$.

\section{Numerical experiments}
\label{sec:3}

We conducted numerical tests for the two cases of synthetic observation data to validate the performance of the proposed method. The assumed true values for the parameters in the Omori-Utsu and the Gutenberg-Richter laws that are used to generate the two synthetic datasets are 
\begin{eqnarray}
b=0.9,\quad \ln K=-3.329, \quad p=1.100, \quad \ln c=-5.809.
\end{eqnarray} 
The magnitude of the main shock is assumed to be $M_0=6.0$. 

The true detection functions modeled by eq.~\eqref{detection} in the two cases are
\begin{enumerate}
\item[Case 1.] $\displaystyle\mu(t)=\frac{5}{1+\exp(15t)}+1.4$;
\item[Case 2.] $\displaystyle\mu(t)=\frac{5}{1+\exp(15t)}+1.4-\sum_{j=1}^4 \nu_j(t)$,
\end{enumerate}
where the same scale parameter defined in eq.~\eqref{detection} is assumed as $s = 0.2$ for both cases,
\begin{eqnarray}
\nu_j(t)=\frac{\sin\llp15(\log_{10}t+1.8-0.4j)\rrp}{j+2}I\llp 15|\log_{10}t+1.8-0.4j|\leq \pi \rrp,
\end{eqnarray}
and $I\{A\}$ is the indicator function that is one if an event $A$ is true, otherwise it is zero. Figure~4 shows the true detection functions for both cases. Case 1 supposes an ordinary situation that the detectable magnitude of aftershocks monotonically decreases as the signal-to-noise ratio recovers with elapsed time $t$ from the main shock. Case 2 supposes a more complicated situation that several large aftershocks excite many subsequent aftershocks, which temporarily decrease the signal-to-noise ratio \citep{Omi2013}. We first estimated the predictive distribution of $\mu(\cdot)$ and $\beta$, and their credibility with the data of aftershocks detected within $t\leq$ 3h, 6h, 12h, and 24h. With the estimated $\hat{\mu}(\cdot)$, $\hat{\beta}$, and $\hat{s}$, $\bm{\tau}$ is estimated by the method stated in Section~2.4. 

\begin{figure}
  \centering
  \includegraphics[width=150mm]{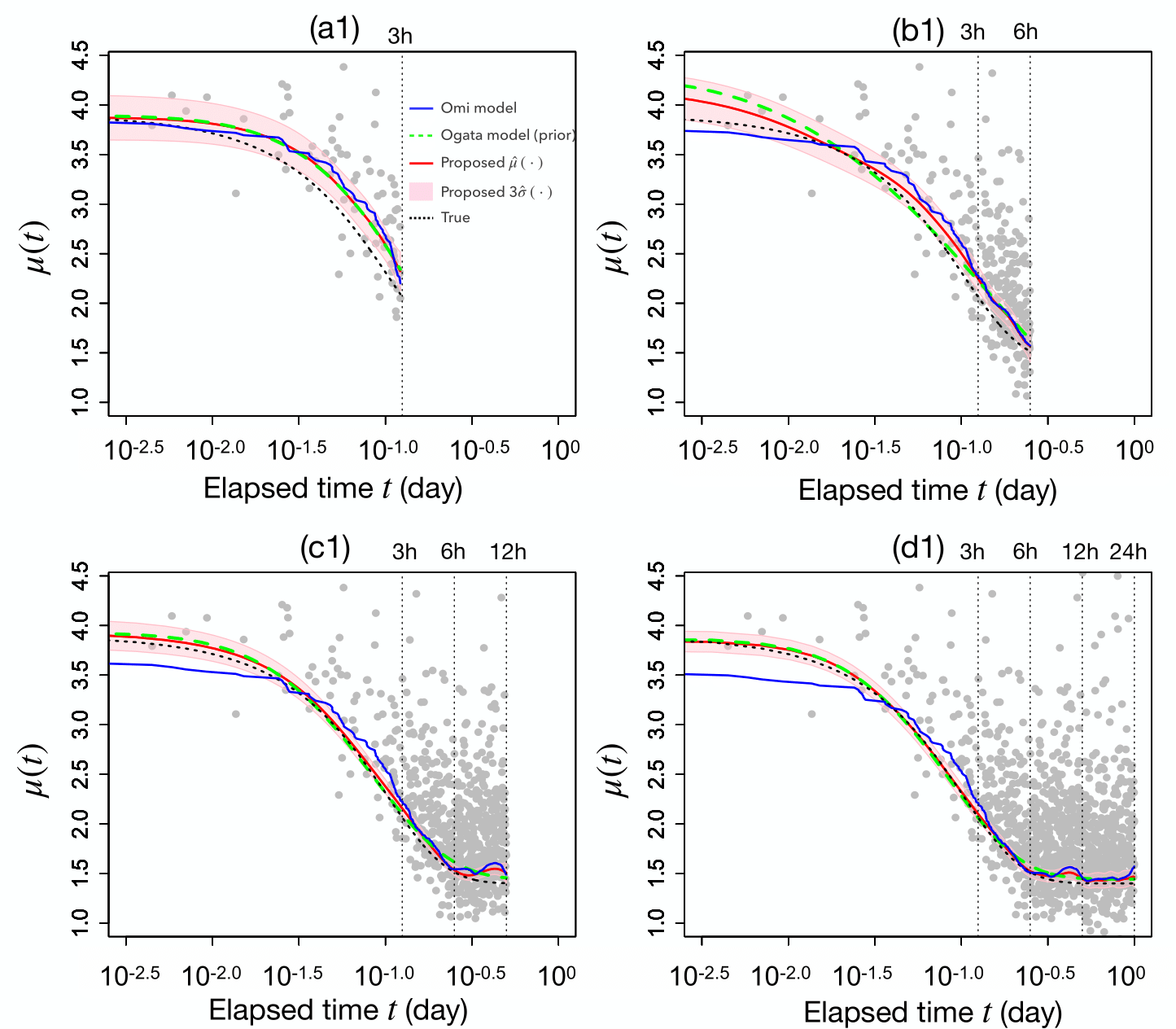}
  \caption{Results of numerical tests to validate the proposed method in Case 1. The blue line indicates the estimated $\mu(\cdot)$ by the method of \citet{Omi2014}, the green line indicates the prior of the mean of the predictive distribution $\mu_{\prior}(\cdot)$ estimated by the method of \citet{Ogata2006}, and the red line indicates the estimated mean $\hat{\mu}(\cdot)$ with the estimation error $\hat{\mu}(\cdot)\pm3\hat{\sigma}(\cdot)$, shown by the red shaded zone, as compared to the true mean $\mu(\cdot)$, shown by the dashed line, assuming that data are available for (a1) $t\leq 3$h, (b1) $t\leq 6$h, (c1) $t\leq 12$h, and (d1) $t\leq 24$h from the main shock.}
  \label{fig:twin}
\end{figure}

Figure~\ref{fig:twin} shows the estimated predictive distribution of $\hat{\mu}(\cdot)$. The mean of the predictive distribution is overall close to the true $\mu(\cdot)$ and inside a region within $3\hat{\sigma}(\cdot)$, where $\hat{\sigma}(\cdot)$ is the estimated standard error of $\hat{\mu}(\cdot)$. In Case 2, it was difficult to detect the first peak in $\mu(\cdot)$ due to the insufficient number of detected aftershocks, but the other peaks could be successfully detected. 
It is worth noting that even if the mean of the prior distribution (green line) is not very similar to the true one (black dashed line), especially in Case 2, the predictive distribution reproduces the true $\mu(\cdot)$ considerably well. The method of \citet{Omi2013} also works well after some time elapses, but it overfits with time when more aftershocks are detected, so that it is biased in time when less aftershocks are detected, such as $t\leq 3$h. Table \ref{newtab:1} shows the $L_2$ distance of between the estimated $\mu(\cdot)$ and the true one:
\begin{eqnarray}
L_2(t)=\int_{0}^{t} (\hat{\mu}(s)-\mu(s))^2ds \label{l2}
\end{eqnarray}
at $t=3/24, 6/24, 12/24$, and $24/24$ (day), which is approximately computed by a left Riemann sum, discretizing the interval $[0,~t]$ into $10^4$ meshes having the same interval. Estimated $\hat{\mu}(\cdot)$ with the Ogata model is the closest to the true one in Case 1, which is an obvious result showing that the Ogata model is appropriate for such a smoothed $\mu(\cdot)$. It is to be noted in Case 1 that the proposed model obtained a more smoothed $\mu(\cdot)$ having a smaller $L_2$ distance compared to that estimated by the Omi method, indicating that the proposed model appropriately avoided overfitting. In Case 2, the Omi and the proposed models, both of which can extract rapid temporal variations in $\mu(\cdot)$, naturally obtained smaller $L_2$ distances than the Ogata model. The remarkable point in Case 2 is that the proposed model obtained the smallest $L_2$ distance among the three models, indicating that the proposed model firmly extracted the temporal variation in the true $\mu(\cdot)$. In summary, the proposed model is as robust as the Ogata model in Case 1 and more flexible than the Omi method for estimating the complicated $\mu(\cdot)$ in Case 2.

Table~\ref{tab:1} summarizes the estimated $\hat{b}$ and $\hat{\bm{\tau}}$.  The resulting $b$ within $t\leq 3$h and 6h in Case 1 overestimates the true value $b=0.9$. This overestimation is not due to the imperfectness of the proposed method but due to the sampling bias. In fact, as the elapsed time increases, the mean of posterior samples converges  to the true value, and the width of the credible interval becomes shorter. 

\newpage

\begin{contfigure}
  \centering
  \includegraphics[width=150mm]{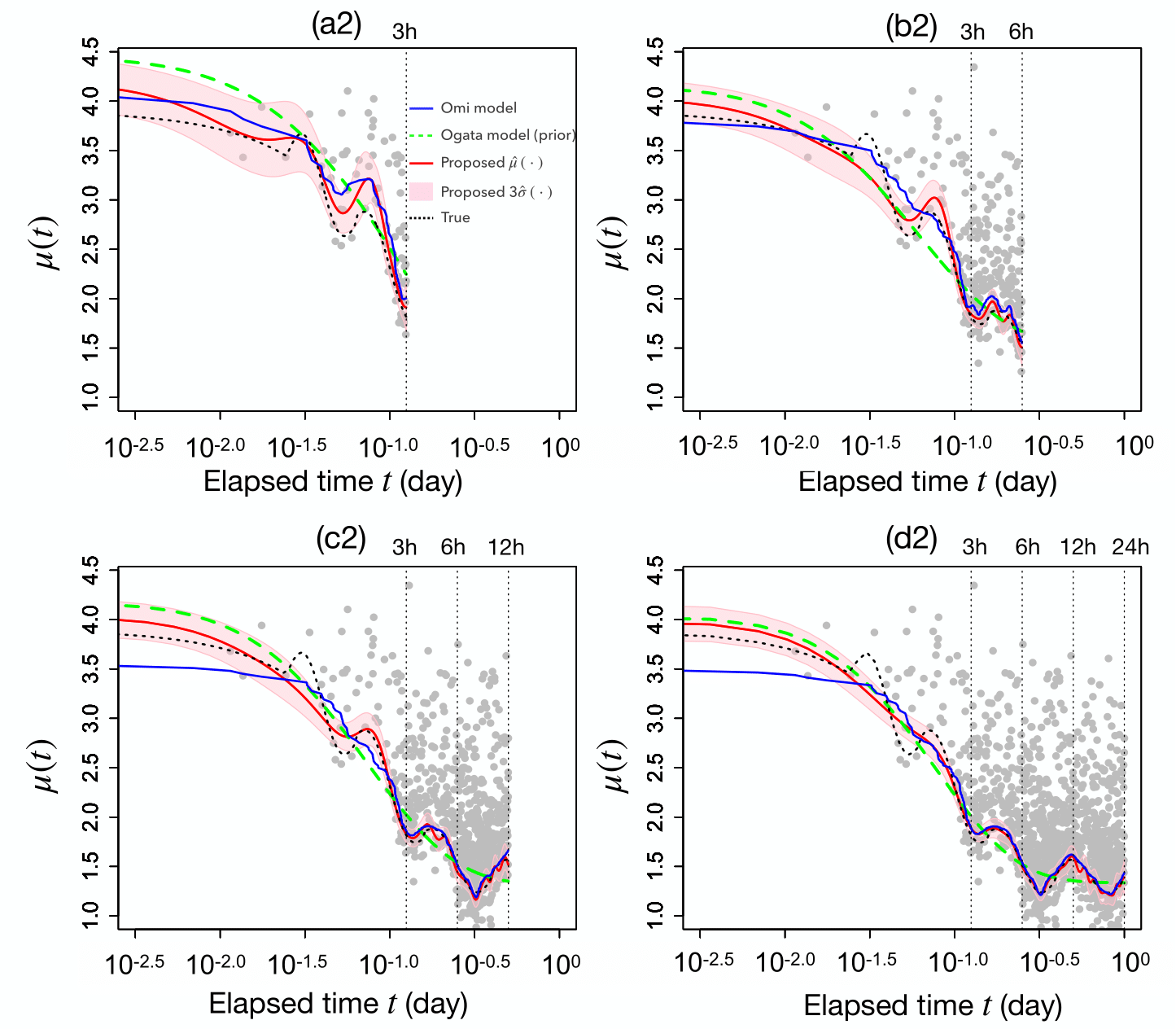}
  \caption{Results of numerical tests to validate the proposed method in Case 2. See the caption of the previous figure for detailed explanations.} 
  \label{fig:twin2}
\end{contfigure}

\begin{table*}
  \caption{$L_2$ distance (eq.~\ref{l2}) of estimated $\mu(\cdot)$ for Ogata, Omi, and proposed methods in Cases 1 and 2. Bold value indicates the closest method to the true $\mu(\cdot)$.}
  \begin{tabular}{cccccccccc} \hline\hline
    &   \multicolumn{4}{c}{Case 1}  & & \multicolumn{4}{c}{Case 2} \\\cline{2-5} \cline{7-10}
    Elapsed time & $t\leq$3h & $t\leq$6h & $t\leq$12h & $t\leq$24h & & $t\leq$3h & $t\leq$6h & $t\leq$12h & $t\leq$24h \\ \hline
     Ogata      & \textbf{0.046} & \textbf{0.016}  & \textbf{0.007} & \textbf{0.001} & & 0.042 & 0.040 & 0.029 & 0.019 \\ 
     Omi         & 0.085 & 0.036  & 0.019 & 0.009 & & 0.082 & 0.019 & 0.011 & 0.007 \\
     Proposed & 0.053 & 0.021  & \textbf{0.007} & 0.002 & &\textbf{0.039} & \textbf{0.013} & \textbf{0.007} & \textbf{0.004} \\  \hline\hline
    \end{tabular}
\label{newtab:1}
\end{table*}

\begin{table*}
  \caption{Estimated posterior means and standard errors of $b$ and $\bm{\tau}=(K, c, p)^\top$ in Cases 1 and 2}
  \begin{tabular}{cccccccccc} \hline\hline
    &   \multicolumn{4}{c}{Case 1}  & & \multicolumn{4}{c}{Case 2} \\\cline{2-5} \cline{7-10}
    Elapsed time & $t\leq$3h & $t\leq$6h & $t\leq$12h & $t\leq$24h & & $t\leq$3h & $t\leq$6h & $t\leq$12h & $t\leq$24h \\ 
    \# of data& 78 & 289 & 680 & 1091 & & 93& 278 & 733 & 1168 \\ \hline
     \multirow{2}{*}{$b$} & 1.060 & 1.035 & 0.974 & 0.936 & & 0.966 & 0.895 & 0.861 & 0.877 \\ 
      & ($\pm$ 0.073) & ($\pm$ 0.056)  & ($\pm$ 0.048) & ($\pm$ 0.031) & & ($\pm$ 0.078) & ($\pm$ 0.062) & ($\pm$ 0.040) & ($\pm$ 0.033) \\ \hline
     \multirow{2}{*}{$\ln K$} & -6.350 & -6.089  & -4.636 & -3.556 & & -5.771 & -4.644 &  -3.472 & -3.024 \\ 
      & ($\pm$ 0.292) & ($\pm$ 0.163)  & ($\pm$ 0.092) & ($\pm$ 0.051) & & ($\pm$ 0.257) & ($\pm$ 0.166) & ($\pm$ 0.085) & ($\pm$ 0.048) \\ 
     \multirow{2}{*}{$p$}       & 1.115 & 1.231  & 1.077 & 1.031 & & 1.084 & 1.068 & 0.986 & 1.047 \\
      & ($\pm$ 0.111) & ($\pm$ 0.086)  & ($\pm$ 0.063) & ($\pm$ 0.044) & & ($\pm$ 0.090) & ($\pm$ 0.082) & ($\pm$ 0.055) & ($\pm$ 0.041) \\ 
     \multirow{2}{*}{$\ln c$} & -5.355 & -5.873  & -5.950 & -5.827 & & -6.823 & -6.348 & -6.603 & -6.229 \\ 
      & ($\pm$ 0.672) & ($\pm$ 0.719)  & ($\pm$ 0.727) & ($\pm$ 0.705) & & ($\pm$ 0.863) & ($\pm$ 0.907) & ($\pm$ 0.896) & ($\pm$ 0.831)  \\   \hline\hline
    \end{tabular}
\label{tab:1}
\end{table*}


\section{Real data analysis}
\label{sec:4}
\begin{figure}
  \centering
  \includegraphics[width=105mm]{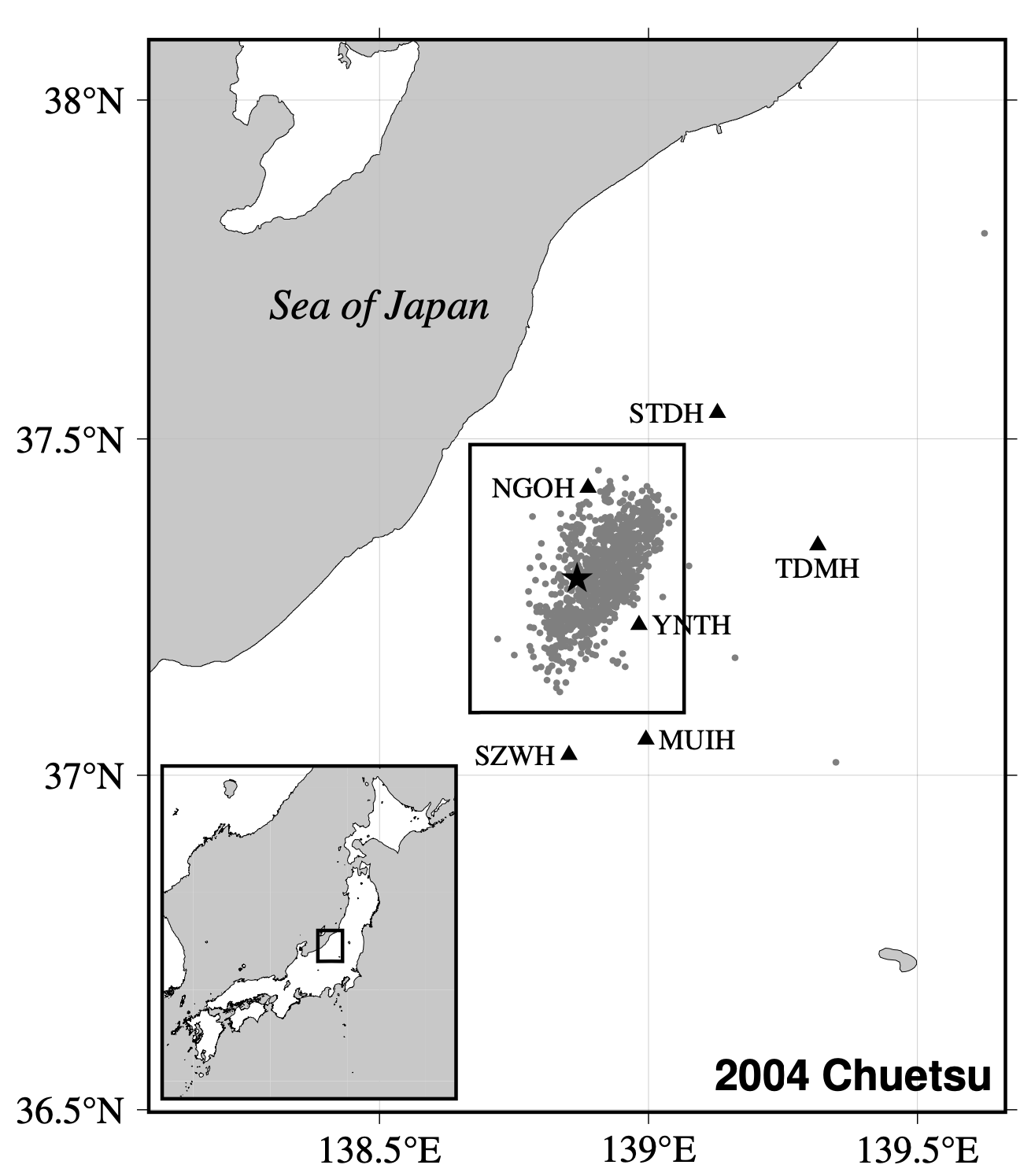}
  \caption{Spatial distribution of aftershocks (circle) within $t\leq 24$h from the main shock (star) of the 2004 Chuetsu earthquake, officially released by JMA. The small rectangle is the Utsu-Seki aftershock zone, the center of which is the epicenter of the main shock, having the angle lengths of $2D(M_0)$ for both latitudinal and longitudinal directions, where $D(M_0)$ is the Utsu-Seki aftershock zone length. The catalog of aftershocks detected at six Hi-net observatories (triangles) proposed by \citet{Enescu2007} was used to validate the proposed method.}
  \label{fig:map}
\end{figure}

The proposed method was applied to the real catalog data related to the 2004 Chuetsu earthquake, officially released from the Japan Meteorological Agency (JMA). The 2004 Chuetsu earthquake (magnitude $M_0=6.8$, epicenter  $37^\circ17'30''$~N,  $138^\circ 52'00''$~E) occurred in Niigata prefecture, Japan at 17h56m on October 23, 2004 (JST).
 Figure~\ref{fig:map} shows the spatial distribution of aftershocks that occurred within $t\leq 24$h from the main shock. The dataset is perfectly the same as used in \citet{Omi2015}, in which the aftershocks occurred in a rectangular area. Lengths four times as long as those of the Utsu-Seki aftershock zone for latitudinal and longitudinal directions were selected. The Utsu-Seki aftershock zone is a rectangular region, in which the epicenter is located at the center, having the angle lengths of $2D(M_0)$ for both latitudinal and longitudinal directions, where $D(M_0)$ is the Utsu-Seki aftershock zone length, defined using the magnitude of the main shock $M_0$ as $D(M_0)=0.01\times10^{0.5M_0-1.8}$  \citep{Utsu1969}. In the case of the 2004 Chuetsu earthquake, $D(M_0)$ is $23' 53''$.

\begin{figure}
  \centering
  \includegraphics[width=150mm]{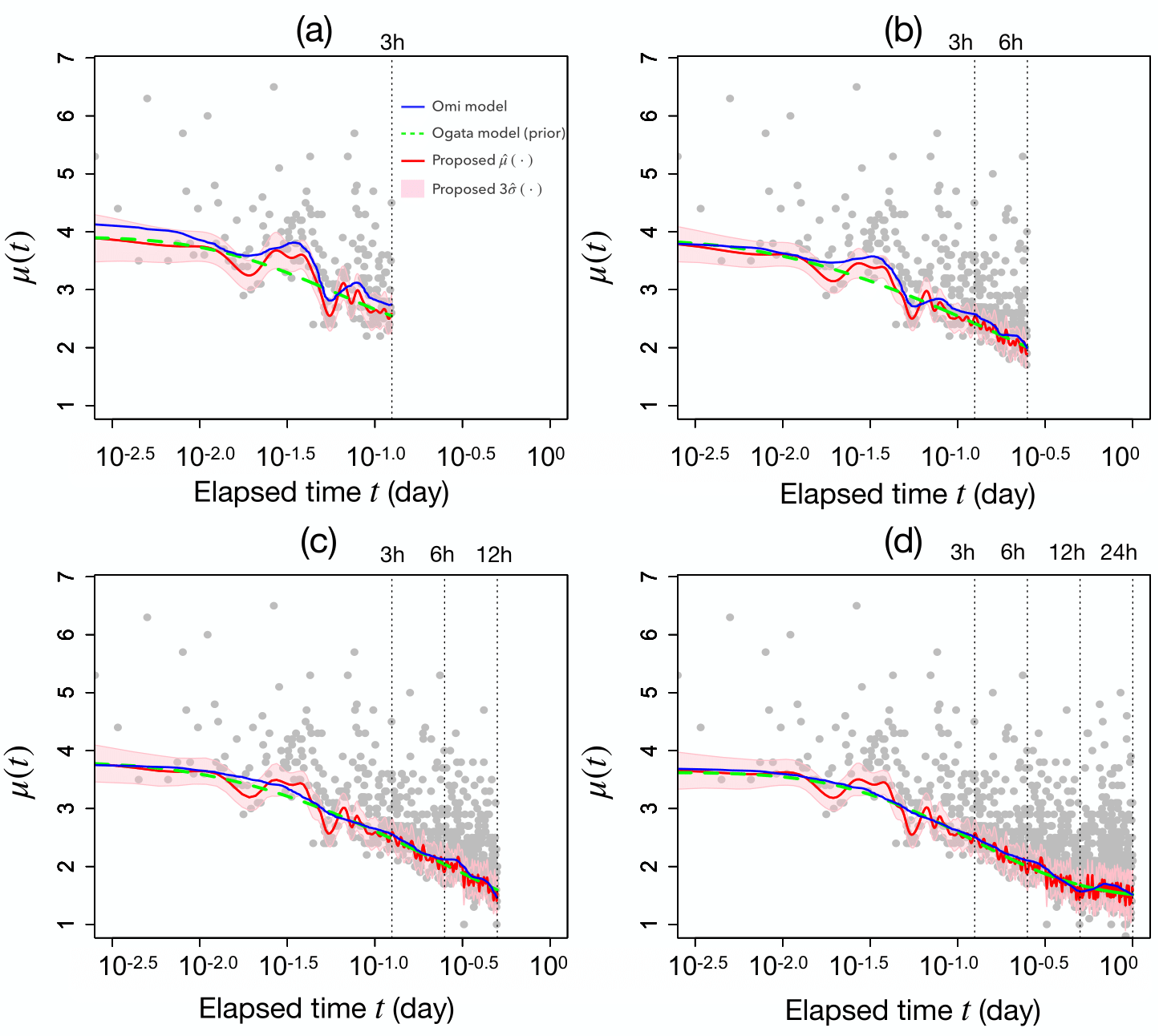}
  \caption{Results of application of our method to the 2004 Chuetsu earthquake based on the JMA catalog. The blue line indicates the estimated $\mu(\cdot)$ by the method of \citet{Omi2014}, the green line indicates the prior of the mean of the predictive distribution $\mu_{\prior}(\cdot)$ estimated by the method of \citet{Ogata2006}, and the red line indicates the estimated mean $\hat{\mu}(\cdot)$ with the estimation error $\hat{\mu}(\cdot)\pm3\hat{\sigma}(\cdot)$, shown by the red shaded zone, assuming that data are available for (a) $t\leq 3$h, (b), $t\leq 6$h (c) $t\leq 12$h, and (d) $t\leq 24$h from the main shock.}
  \label{fig:real1}
\end{figure}

Figure~\ref{fig:real1} shows the results of the application to the real catalog data. Following the same procedure as in the numerical tests in Section 3, the proposed method estimates the mean $\hat{\mu}(\cdot)$ of the predictive distribution with the standard error $\hat{\sigma}(\cdot)$ starting from the prior $\mu_{\prior}(\cdot)$, assuming data availability for $t\leq $3h, 6h, 12h, and 24h from the main shock. The magnitudes of aftershocks range from 0.8 to 6.6 with elapsed time within $t\leq 24$h.

\begin{figure}
  \centering
  \includegraphics[width=130mm]{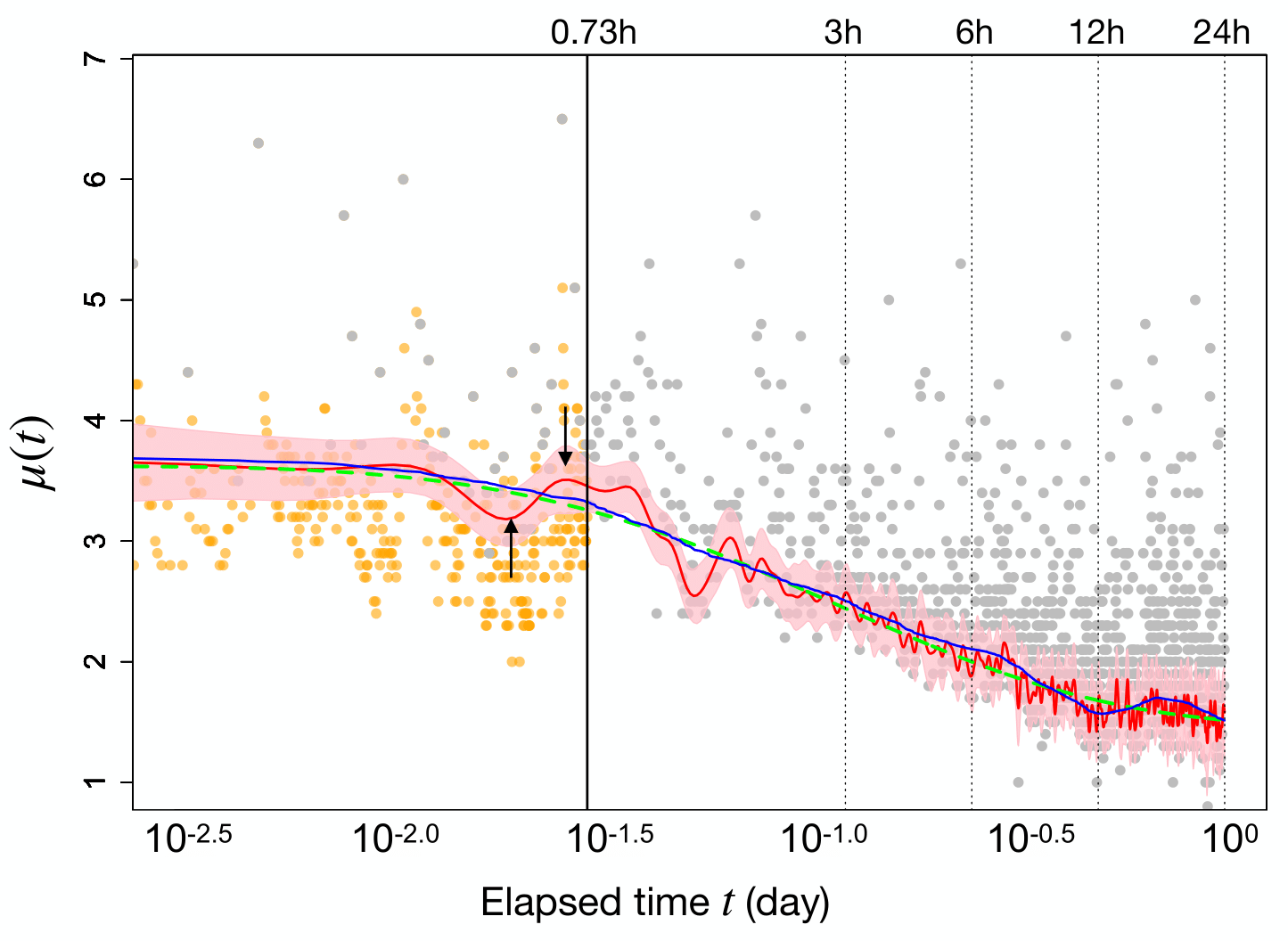}
  \caption{Visually detected aftershocks before $t\leq 0.73$h (orange dots) proposed by \citet{Enescu2007} are added to Figure~\ref{fig:real1}(d). Two arrows indicate the differences between $\mu(t)$ estimated by the proposed method and the method of \citet{Omi2014}. See the caption of Figure~\ref{fig:real1} for detailed explanations.}
  \label{fig:real2}
\end{figure}

An enormous number of large aftershocks would cause and self-excite a sudden decrease of the detection rate. The proposed method can extract such a hidden structure owing to the wide representability of functions of the GPR. 
The estimated $\mu(\cdot)$ by the method of \citet{Omi2013} shown in Figure~\ref{fig:real1} for comparison seems to be unstable, depending on the number of detected aftershocks. When available data are insufficient (Figures \ref{fig:real1}(a) and \ref{fig:real1}(b)), the method of \citet{Omi2013} also captures some oscillations, but after time elapses (Figures \ref{fig:real1}(c) and \ref{fig:real1}(d)), the resulting detection function becomes too smooth. A similar phenomenon is also seen in Figure~3 in \citet{Omi2013}. On the other hand, the proposed method always obtains a stable estimation, independent of the number of aftershocks owing to the wide representability of functions of the GPR, as stated above. To investigate the large difference in the estimates of $\mu(t)$ at $t\leq 1$h, we utilized the catalog of aftershocks proposed by \citet{Enescu2007}. This catalog contains the aftershocks visually detected within $t\leq 0.73$h from six High Sensitivity Seismograph Network (Hi-net) stations, shown in Figure~\ref{fig:map}, which are not included in the JMA catalog. Figure~\ref{fig:real2}, in which the visually detected aftershocks are added to Figure~\ref{fig:real1}(d), supports the validity of the negative peak at $t=0.47$h and the positive peak at $t=0.65$h in $\mu(t)$ estimated by the proposed method. The negative peak reflects the fact that small aftershocks down to $M=2.0$ are detected probably due to a temporal decrease in seismic activities, and the positive peak reflects the increase in the lower bound of observed magnitudes.

\begin{figure}
  \centering
  \includegraphics[width=155mm]{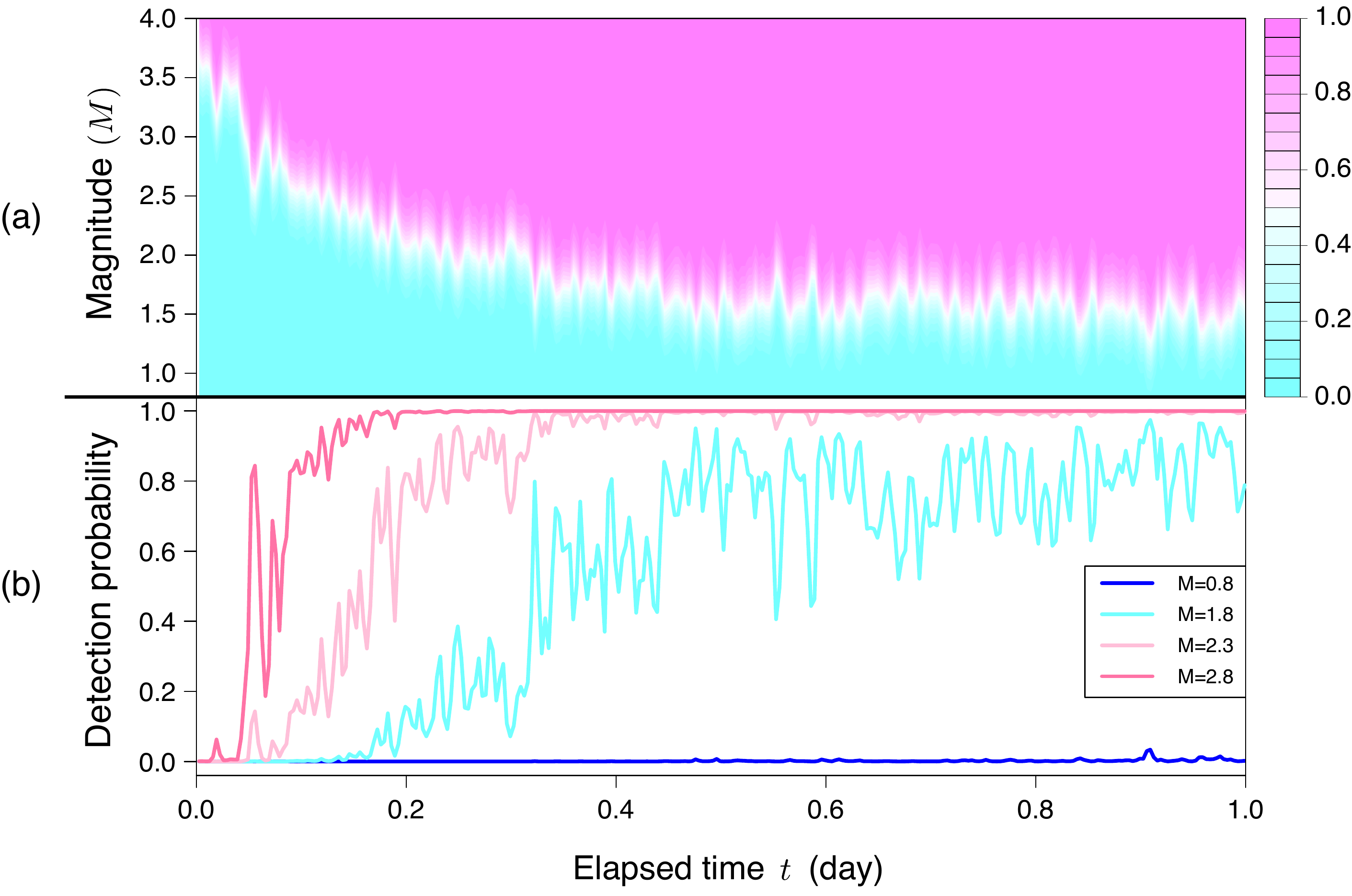}
  \caption{Estimated predictive distribution of detection function $\pi^*(M_1^*, t_1^*)$ with the application of the proposed method to the catalog data related to the 2004 Chuetsu earthquake: (a) predictive distribution of detection probability; (b) cross-sections of the predictive distribution of detection probability at the magnitudes $M=$ 0.8,\;1.8,\;2.3,\;2.8.}
  \label{fig:real3}
\end{figure}

Figure~\ref{fig:real3}(a) shows the estimated predictive detection function (eq. \ref{pred_detect}) in the case of the 2004 Chuetsu earthquake, and Figure~\ref{fig:real3}(b) plots its cross-sections at the magnitudes $M=0.8$, $1.8$, $2.3$, and $2.8$. These magnitudes correspond to the minimum ($0$th percentile), $25$th percentile, median ($50$th percentile), and $75$th percentile in the catalog data, where the $q$th-percentile is the magnitude below which $q\%$ of all the aftershocks are found.
Figure~\ref{fig:real3}(a) shows that the detection function intensively fluctuates, unlike in the numerical experiments, which is thought to reflect the rapid temporal variation of the seismic activities.
Figure~\ref{fig:real3}(b) indicates that the minimum magnitude of the complete recording $M_c$ is 2.3 (75th percentile) after $t\geq 0.4$ (day). \citet{Wiemer2000} proposed a method to estimate $M_c$ by the smallest magnitude at which 90\% of the aftershocks can be modeled by the Gutenberg-Richter law. \citet{Enescu2007} obtained $M_c=1.8$ at $t=0.5$ (day) by applying the procedure of \citet{Wiemer2000} to the 2004 Chuetsu Earthquake, which seems to be consistent with Figure~\ref{fig:real3}(b). Table~\ref{tab:2} summarizes the mean and the standard error of the posterior distribution of $b$ within an elapsed time of $t\leq 3$h, 6h, 12h, and 24h. The estimates for the $b$ are stable, irrespective of the time available, but the estimated $p$ at $t\leq 12$h is somewhat larger compared to the other times. This indicates limitations of the Omori-Utsu law, and it would be better to apply more realistic and complex models, such as the ETAS model \citep{Ogata1988_2}, for the distribution of elapsed times of the aftershocks.

\begin{figure}
  \centering
  \includegraphics[width=155mm]{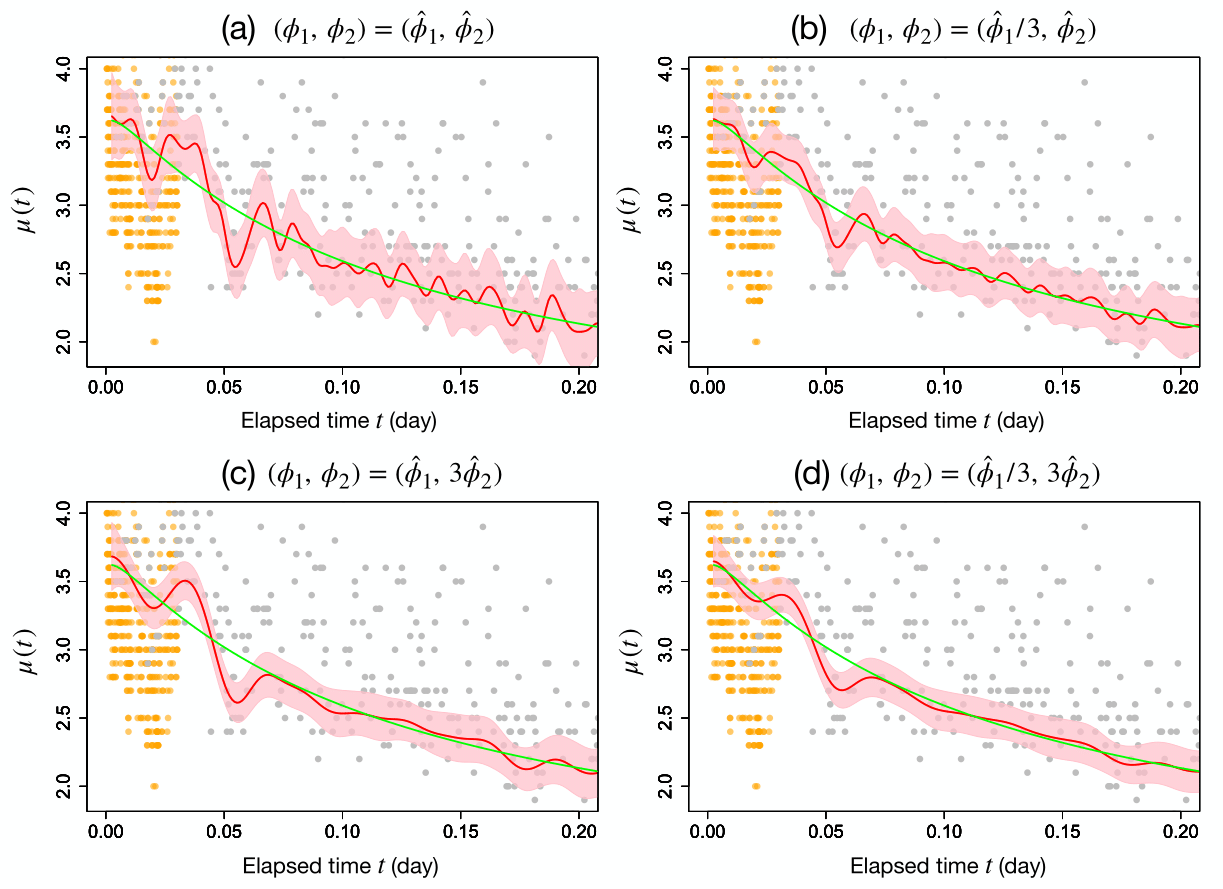}
  \caption{Predictive distribution of $\mu(\cdot)$ with different $(\phi_1, \phi_2)$: (a) $(\hat{\phi}_1, \hat{\phi}_2)$; (b) $(\hat{\phi}_1/3, \hat{\phi}_2)$; (c) $(\hat{\phi}_1, 3\hat{\phi}_2)$; (d) $(\hat{\phi}_1/3, 3\hat{\phi}_2)$, where $(\hat{\phi}_1, \hat{\phi}_2)$ is the estimated hyperparameters.}
  \label{fig:new3}
\end{figure}

Here, we conduct an additional numerical check to show that the estimated $\mu(\cdot)$ has not been affected by overfitting. Let the estimated hyperparameters for $\phi_1$ and $\phi_2$ be $\hat{\phi}_1=0.027$ and $\hat{\phi}_2=0.004$, respectively. The radial basis kernel function (eq.~\ref{kernel}) indicates that small $\phi_1$ and large $\phi_2$ lead to a smoothed $\mu(\cdot)$ function. We re-estimate $b$ and $\bm{\tau}$ with three additional cases with fixed $(\phi_1, \phi_2)$: $(\hat{\phi}_1/3, \hat{\phi}_2)$,  $(\hat{\phi}_1, 3\hat{\phi}_2)$, and $(\hat{\phi}_1/3, 3\hat{\phi}_2)$. Figure \ref{fig:new3} shows predictive distributions of $\mu(\cdot)$ with the estimated hyperparameters (Figure \ref{fig:new3}a) and the three different cases (Figures \ref{fig:new3}b-d). The estimated $\mu(\cdot)$ oscillates the most in Figure~\ref{fig:new3}(a) and the least in Figure~\ref{fig:new3}(d). The most notable point in Figure~\ref{fig:new3} is that the negative peak at $t = 0.47$h is robustly extracted regardless of the fluctuations in the hyperparameters $\phi_1$ and $\phi_2$. This result is considered to indicate that the proposed method properly extracts the temporal variation in $\mu(\cdot)$, without being affected by overfitting.

\begin{table}
  \caption{Estimated posterior means and standard errors of $b$ and $\bm{\tau}=(K, p, c)^\top$ in the case of the 2004 Chuetsu earthquake. }
  \begin{tabular}{ccccc} \hline\hline
   Elapsed time & $t\leq$3h & $t\leq$6h & $t\leq$12h & $t\leq$24h\\ 
   \# of data & 192 & 355 & 655 & 1099 \\ \hline
    \multirow{2}{*}{$b$} & 0.836 & 0.741 & 0.809 & 0.791\\ 
     & ($\pm$ 0.056) & ($\pm$ 0.044)  & ($\pm$ 0.035) & ($\pm$ 0.028)\\ 
    \multirow{2}{*}{$\ln K$} & -5.388 & -4.086  & -4.311 & -3.230\\ 
     & ($\pm$ 0.262) & ($\pm$ 0.173)  & ($\pm$ 0.093) & ($\pm$ 0.049)\\ 
    \multirow{2}{*}{$p$} & 1.150 & 1.201  & 1.342 & 1.228\\ 
     & ($\pm$ 0.110) & ($\pm$ 0.097)  & ($\pm$ 0.073) & ($\pm$ 0.049)\\ 
    \multirow{2}{*}{$\ln c$} & -4.671 & -4.358  & -4.304 & -4.498\\ 
     & ($\pm$ 0.481) & ($\pm$ 0.415)  & ($\pm$ 0.341) & ($\pm$ 0.345)\\ \hline\hline
     \end{tabular}
\label{tab:2}
\end{table}

\section{Concluding remarks}
\label{sec:5}

Immediate prediction of seismic activities after the main shock is important to assess hazards of subsequent aftershocks. Contaminations of arriving seismic waves right after the main shock interfere with counting the number of aftershocks correctly, so that the number of detected aftershocks is underestimated. 
 This underestimated count causes distorted estimates for the distribution of aftershocks or seismic activities.
To rapidly and stably estimate the temporal changes in the occurrence rate of aftershocks with limited data right after a main shock, we introduced a GPR-based detection function to remove the effects of undetected aftershocks.
Owing to the nonparametric and Bayesian properties in the GPR, the proposed detection function has four advantages superior to previous methods: (i)~the resulting estimates are stable by virtue of adoption of prior information; (ii)~specification of the detection function is semiparametric in the sense that $\pi(t, M; \mu, s)$ is parametric, but $\mu$ is nonparametric; (iii)~MCMC sampling is effective in computing hyperparameters without computation of complicated integrations; (iv)~credible intervals can be obtained in a natural way.

The limitation of the proposed method lies in the assumption on the joint intensity function (eq.~\ref{joint}). It is known that real catalog data should be described by more complex intensity functions, such as represented by the ETAS model \citep{Ogata1988_2}. The proposed method can be extended straightforwardly to the ETAS model, as done by \citet{Omi2014}, which remains as future work. 

\begin{acknowledgments}
This work was supported by the Tokyo Metropolitan Resilience Project of the National Research Institute for Earth Science and Disaster Resilience (NIED), and JST CREST Grant Numbers JPMJCR1763 and JPMJCR1761. The key ideas in this study came through the activities of JSPS KAKENHI Grant-in-Aids for Early-Career Scientists No. 19K14592, 19K14671, 20K19756, Grant-in-Aids for Scientific Research (B) No. 17H01703, 17H01704, 18H03210, and Grant-in-Aid for Scientific Research (S) No. 19H05662. The travel expenses needed to discuss among co-authors was partially supported by ERI JURP 2020-A-05. One of the figures was drawn using the software Generic Mapping Tools (GMT)  developed by \citet{Wessel1998}. We appreciate Prof. Aitaro Kato for constructive discussions and Prof. Bogdan Enescu for providing us his catalog related to the 2004 Chuetsu earthquake.
\end{acknowledgments}

\section*{DATA AVAILABILITY}
The official catalog of events associated with the 2004 Chuetsu earthquake is available via the website of the JMA (\url{http://www.data.jma.go.jp/svd/eqev/data/bulletin/}). Another catalog that also contained visually detected aftershocks \citep{Enescu2007} was provided by Prof. Bogdan Enescu.

\bibliographystyle{gji2}
\bibliography{refs}

\begin{thebibliography}{40}
\expandafter\ifx\csname natexlab\endcsname\relax\def\natexlab#1{#1}\fi

\bibitem[Aki(1965)]{Aki1965}
Aki, K., 1965.
\newblock Maximum likelihood estimate of $b$ in the formula $\log {N}=a-b{M}$
  and its confidence limits, {\it Bull. Earthq. Res. Inst. Univ. Tokyo\/}, {\bf
  43}, 237--239.

\bibitem[de~G.~Matthews et~al.(2018)de~G.~Matthews, Rowland, Hron, Turner, \&
  Ghahramani]{Matthews2018}
de~G.~Matthews, A.~G., Rowland, M., Hron, J., Turner, R.~E., \& Ghahramani, Z.,
  2018.
\newblock Gaussian process behaviour in wide deep neural networks, {\it arXiv
  preprint: 1804.11271\/}.

\bibitem[Enescu et~al.(2007)Enescu, Mori, \& Miyazawa]{Enescu2007}
Enescu, B., Mori, J., \& Miyazawa, M., 2007.
\newblock Quantifying early aftershock activity of the 2004 mid-{N}iigata
  {P}refecture earthquake (${M}_w$6.6), {\it J. Geophys. Res. Solid Earth\/},
  {\bf 112}, 859--869.

\bibitem[Genz \& Bretz(2009)]{Genz2009}
Genz, A. \& Bretz, F., 2009.
\newblock {\it Computation of Multivariate Normal and $t$ Probabilities.
  Lecture Notes in Statistics.\/}, Heidelberg: Springer.

\bibitem[Geweke(1991)]{Geweke1991}
Geweke, J.~F., 1991.
\newblock Efficient simulation from the multivariate normal and student-$t$
  distributions subject to linear constraints and the evaluation of constraint
  probabilities, {\it \rm Computer Science and Statistics. Proceedings of the
  23rd Symposium on the Interface. Seattle Washington, April 21-24, 1991,
  571--578\/}.

\bibitem[Geweke(2005)]{Geweke2005}
Geweke, J.~F., 2005.
\newblock {\it Contemporary Bayesian Econometrics and Statistics\/}, Wiley \&
  Sons.

\bibitem[Gutenberg \& Richter(1944)]{Gutenberg1944}
Gutenberg, B. \& Richter, C.~F., 1944.
\newblock Frequency of earthquakes in {California}, {\it Bull. Seism. Soc.
  Am.\/}, {\bf 34}, 185--188.

\bibitem[Hainzl(2016)]{Hainzl2016}
Hainzl, S., 2016.
\newblock Rate-dependent incompleteness of earthquake catalogs, {\it Seismol.
  Res. Lett.\/}, {\bf 87}, 337--344.

\bibitem[Kagan \& Jackson(2000)]{Kagan2000}
Kagan, Y.~Y. \& Jackson, D.~D., 2000.
\newblock Probabilistic forecasting of earthquakes, {\it Geophys. J. Int.\/},
  {\bf 143}, 438--453.

\bibitem[Kijko \& Smit(2017)]{Kijko2017}
Kijko, A. \& Smit, A., 2017.
\newblock Estimation of the frequency-magnitude gutenberg-richter $b$-value
  without making assumptions on levels of completeness, {\it Seismol. Res.
  Lett.\/}, {\bf 88}, 311--318.

\bibitem[Kuwatani et~al.(2018)Kuwatani, Nagao, Ito, Okamoto, Yoshida, \&
  Okudaira]{Kuwatani2018}
Kuwatani, T., Nagao, H., Ito, S., Okamoto, A., Yoshida, K., \& Okudaira, T.,
  2018.
\newblock Recovering the past history of natural recording media by {Bayesian}
  inversion, {\it Phys. Rev. E\/}, {\bf 98}, 043311.

\bibitem[Martinsson \& Jonsson(2018)]{Martinsson2018}
Martinsson, J. \& Jonsson, A., 2018.
\newblock A new model for the distribution of observable earthquake magnitudes
  and applications to $b$-value estimation, {\it IEEE Geosci. Rem. Sens.
  Lett.\/}, {\bf 15}, 833--837.

\bibitem[Metropolis et~al.(1953)Metropolis, Rosenbluth, Rosenbluth, \&
  Teller]{Metropolis1953}
Metropolis, N., Rosenbluth, A.~W., Rosenbluth, M.~N., \& Teller, E., 1953.
\newblock Equations of state calculations by fast computing machine, {\it J.
  Chem. Phys.\/}, {\bf 21}, 1087--1091.

\bibitem[Mignan(2012)]{Mignan2012}
Mignan, A., 2012.
\newblock Functional shape of the earthquake frequency‐magnitude distribution
  and completeness magnitude, {\it J. Geophys. Res. Solid Earth\/}, {\bf 117},
  B08302.

\bibitem[Mignan(2019)]{Mignan2019}
Mignan, A., 2019.
\newblock Generalized earthquake frequency–magnitude distribution described
  by asymmetric laplace mixture modelling, {\it Geophys. J. Int.\/}, {\bf 219},
  1348--1364.

\bibitem[Ogata(1983)]{Ogata1983}
Ogata, Y., 1983.
\newblock Estimation of the parameters in the modified omori formula for
  aftershock frequencies by the maximum likelihood procedure, {\it J. Phys.
  Earth\/}, {\bf 31}, 115--124.

\bibitem[Ogata(1988)]{Ogata1988_2}
Ogata, Y., 1988.
\newblock Statistical models for earthquake occurrences and residual analysis
  for point processes, {\it J. Amer. Statist. Assoc.\/}, {\bf 83}, 29--39.

\bibitem[Ogata \& Katsura(1993)]{Ogata1993}
Ogata, Y. \& Katsura, K., 1993.
\newblock Analysis of temporal and spatial heterogeneity of magnitude frequency
  distribution inferred from earthquake catalogs, {\it Geophys. J. Int.\/},
  {\bf 113}, 727--738.

\bibitem[Ogata \& Katsura(2006)]{Ogata2006}
Ogata, Y. \& Katsura, K., 2006.
\newblock Immediate and updated forecasting of aftershock hazard, {\it Geophys.
  Res. Lett.\/}, {\bf 33}, L10305.

\bibitem[Omi et~al.(2013)Omi, Ogata, Hirata, \& Aihara]{Omi2013}
Omi, T., Ogata, Y., Hirata, Y., \& Aihara, K., 2013.
\newblock Forecasting large aftershocks within one day after the main shock,
  {\it Sci. Rep.\/}, {\bf 3}, 2218.

\bibitem[Omi et~al.(2014)Omi, Ogata, Hirata, \& Aihara]{Omi2014}
Omi, T., Ogata, Y., Hirata, Y., \& Aihara, K., 2014.
\newblock Estimating the {ETAS} model from an early aftershock sequence, {\it
  Geophys. Res. Lett.\/}, {\bf 41}, 850--857.

\bibitem[Omi et~al.(2015{\natexlab{a}})Omi, Ogata, Hirata, \&
  Aihara]{Omi2015_2}
Omi, T., Ogata, Y., Hirata, Y., \& Aihara, K., 2015{\natexlab{a}}.
\newblock Intermediate-term forecasting of aftershocks from an early aftershock
  sequence: Bayesian and ensemble forecasting approaches, {\it J. Geophys. Res.
  Solid Earth\/}, {\bf 120}, 2561--2578.

\bibitem[Omi et~al.(2015{\natexlab{b}})Omi, Ogata, Shiomi, Enescu, Sawazaki, \&
  Aihara]{Omi2015}
Omi, T., Ogata, Y., Shiomi, K., Enescu, B., Sawazaki, K., \& Aihara, K.,
  2015{\natexlab{b}}.
\newblock Automatic aftershock forecasting: A test using real-time seismicity
  data in {Japan}, {\it Bull. Seismol. Soc. Am.\/}, {\bf 106}, 2450--2458.

\bibitem[Omori(1894)]{Omori1894}
Omori, F., 1894.
\newblock On the aftershocks of earthquake, {\it J. ColI. Sci. Imp. Univ.
  Tokyo\/}, {\bf 7}, 111--200.

\bibitem[Petersen \& Petersen(2012)]{Petersen2012}
Petersen, K.~B. \& Petersen, M.~S., 2012.
\newblock The matrix cookbook, {\it \rm Technical report, Technical University
  of Denmark, 2007. URL http://www2.imm.dtu.dk/pubdb/p.php?3274\/}.

\bibitem[Qin(2017)]{Qin2017}
Qin, J., 2017.
\newblock {\it Biased Sampling, Over-identified Parameter Problems and
  Beyond\/}, Singapore: Springer.

\bibitem[Rasmussen et~al.(2006)Rasmussen, Williams, \&
  Christopher]{Rasmussen2006}
Rasmussen, C.~E., Williams, C. K.~I., \& Christopher, K.~I., 2006.
\newblock {\it Gaussian Processes for Machine Learning\/}, The MIT Press.

\bibitem[Resenberg \& Jones(1989)]{Resenberg1989}
Resenberg, P.~A. \& Jones, L.~M., 1989.
\newblock Earthquake hazard after a mainshock in {California}, {\it Science\/},
  {\bf 243}, 1173--1176.

\bibitem[Resenberg \& Jones(1994)]{Resenberg1994}
Resenberg, P.~A. \& Jones, L.~M., 1994.
\newblock Earthquake aftershocks: {Update}, {\it Science\/}, {\bf 265},
  1251--1252.

\bibitem[Ringdal(1975)]{Ringdal1975}
Ringdal, F., 1975.
\newblock On the estimation of seismic detection thresholds, {\it Bull. Seism.
  Soc. Am.\/}, {\bf 65}, 1631--1642.

\bibitem[Tanner \& Wong(1987)]{Tanner1987}
Tanner, M.~A. \& Wong, W.~H., 1987.
\newblock The calculation of posterior distributions by data augmentation, {\it
  J. Am. Stat. Assoc.\/}, {\bf 82}, 528--540.

\bibitem[Utsu(1961)]{Utsu1961}
Utsu, T., 1961.
\newblock A statistical study on the occurrence of aftershocks, {\it J. ColI.
  Sci. Imp. Univ. Tokyo\/}, {\bf 30}, 521--605.

\bibitem[Utsu(1969)]{Utsu1969}
Utsu, T., 1969.
\newblock Aftershocks and earthquake statistics ({I}) --- {Some} parameters
  which characterize an aftershock sequence and their interrelations ---, {\it
  J. Fac. Sci., Hokkaido Univ., Ser. VII (Geophysics)\/}, {\bf 3}, 129--195.

\bibitem[Utsu(1970)]{Utsu1970}
Utsu, T., 1970.
\newblock Aftershocks and earthquake statistics ({II}) --- {Further}
  investigation of aftershocks and other earthquake sequences based on a new
  classification of earthquake sequences ---, {\it J. Fac. Sci. Hokkaido Univ.,
  Ser. VII (Geophysics)\/}, {\bf 3}, 197--266.

\bibitem[Vardi(1982)]{Vardi1982}
Vardi, Y., 1982.
\newblock Nonparametric estimation in presence of length bias, {\it Ann.
  Stat.\/}, {\bf 10}, 616--620.

\bibitem[Vardi(1985)]{Vardi1985}
Vardi, Y., 1985.
\newblock Empirical distributions in selection bias models, {\it Ann. Stat.\/},
  {\bf 13}, 178--203.

\bibitem[Wessel \& Smith(1998)]{Wessel1998}
Wessel, P. \& Smith, W. H.~F., 1998.
\newblock New, improved version of generic mapping tools released, {\it Trans.
  Am. Geophys. Un.\/}, {\bf 79}, 579.

\bibitem[Wiemer \& Wyss(2000)]{Wiemer2000}
Wiemer, S. \& Wyss, M., 2000.
\newblock Minimum magnitude of completeness in earthquake catalogs: Examples
  from alaska, the western united states, and japan, {\it Bull. Seismol. Soc.
  Am.\/}, {\bf 90}, B04310.

\bibitem[Wilhelm \& Manjunath(2015)]{tmvtnorm}
Wilhelm, S. \& Manjunath, B.~G., 2015.
\newblock {\it {tmvtnorm}: Truncated Multivariate Normal and Student t
  Distribution\/}, R package version 1.4-10.

\bibitem[Zhuang et~al.(2017)Zhuang, Ogata, \& Wang]{Zhuang2017}
Zhuang, J., Ogata, Y., \& Wang, T., 2017.
\newblock Data completeness of the kumamoto earthquake sequence in the jma
  catalog and its influence on the estimation of the etas parameters, {\it
  Earth, Planets and Space\/}, {\bf 69}, 36.

\end{thebibliography}

\appendix
\section{Theoretical results}
\begin{lemma}
\textbf{(Formula of sum of two squared forms \citep[][section 8.1.7]{Petersen2012})}.
\end{lemma}

For any vectors $\bm{m}_1$ and $\bm{m}_2$, and nonsingular matrices $\Sigma_1$ and $\Sigma_2$, it holds that
\begin{eqnarray}
&&-\frac{1}{2}(\bm{x}-\bm{m}_1)^\top \Sigma^{-1}_1(\bm{x}-\bm{m}_1)\nonumber\\
&&\quad -\frac{1}{2}(\bm{x}-\bm{m}_2)^\top \Sigma^{-1}_2(\bm{x}-\bm{m}_2)\nonumber\\
&&=-\frac{1}{2}(\bm{x}-\bm{m}_c)^\top \Sigma^{-1}_c(\bm{x}-\bm{m}_c)+C,\label{formula}
\end{eqnarray}
where
\begin{eqnarray}
\Sigma_c^{-1} &=& \Sigma_1^{-1}+\Sigma_2^{-1}\\
\bm{m}_c &=& (\Sigma_1^{-1}+\Sigma_2^{-1})^{-1}(\Sigma_1^{-1}\bm{m}_1+\Sigma_2^{-1}\bm{m}_2)\\
C &=& \frac{1}{2}\bm{m}_c^\top \Sigma^{-1}_c \bm{m}_c-\frac{1}{2}(\bm{m}_1^\top \Sigma_1^{-1}\bm{m}_1+\bm{m}_2^\top \Sigma_2^{-1}\bm{m}_2).
\end{eqnarray}

\vspace{2ex}
\noindent
\textbf{Proof of Eq.~\eqref{marginal}.} Because $p(\beta)$ does not involve $\bm{\mu}$, the definition of the posterior distribution is
\begin{eqnarray}
L(\bm{\theta}) = p(\beta)\int p(\bm{M}_{1}\mid  \bm{\mu}; \beta, s^2) p(\bm{\mu}\mid \bm{\mu}_{\prior}, \mK_n)d\bm{\mu},
\end{eqnarray}
where $\theta=(\beta, s^2, \phi^2_2)^\top$, $\bm{\mu}_{\prior}=(\mu_{\prior}(t_{11}), \ldots, \mu_{\prior}(t_{1n}))^\top$ is the mean of the prior distribution, and $\mK_n=(\mK(t_{1i},t_{1j}))$ is the variance of the prior distribution, which is an $n$ by $n$ matrix. Hereafter, we ignore $p(\beta)$ because it does not have an effect on integration. Recall that the definition of the conditional distribution of $M_1$ given $\mu$ is given in eq.~\eqref{detect_data}. It follows from
\begin{eqnarray}
&&\exp(\beta \mu_i) \Phi(M_{1i}; \mu_i, s^2)\nonumber\\
&&= \frac{\exp(\beta \mu_i)}{\sqrt{2\pi s^2}} \int_{-\infty}^{M_{1i}} \exp\llp-\frac{(x-\mu_i)^2}{2s^2}\rrp dx\nonumber\\
&&= \frac{\exp(s^2 \beta^2/2)}{\sqrt{2\pi s^2}} \int_{-\infty}^{M_{1i}} \exp\lllp \beta x-\frac{\{\mu_i-(x+s^2\beta)\}^2}{2s^2}\rrrp dx
\end{eqnarray}
 that the conditional distribution can be rewritten as
\begin{eqnarray}
&&\prod_{i=1}^n p(M_{1i}\mid  \mu_i; \beta, s^2)\nonumber\\
&&=\beta^n \exp\llp-\beta \sum_{i=1}^n M_{1i}-\frac{n}{2}\beta^2 s^2\rrp\prod_{i=1}^n \exp(\beta \mu_i)\Phi(M_{1i}; \mu_i, s^2)\nonumber\\
&&
\begin{split}=& \frac{\beta^n \exp(-\beta \sum_{i=1}^n M_{1i})}{\sqrt{(2\pi)^n |\Sigma|}}\int_{\bm{x}\in \mM} \exp(\beta \bm{1}^\top \bm{x} )\\
&\quad \times \exp\lllp -\frac{1}{2}\{\bm{\mu}-(\bm{x}+\beta\Sigma \bm{1})\}^\top \Sigma^{-1}\{\bm{\mu}-(\bm{x}+\beta\Sigma \bm{1})\}\rrrp d\bm{x},
\end{split}
\end{eqnarray}
where $\Sigma=s^2 I_n$, $I_n$ is a $n$ by $n$ identity matrix, and $\mM=\otimes_{i=1}^n \{x_i \leq M_{1i}\}$. Letting $\bm{m}_1=\bm{x}+\beta \Sigma \bm{1}$, $\bm{m}_2=\bm{\mu}_{\prior}$, $\Sigma_1=\Sigma$, and $\Sigma_2=\mK_n$ with the formula (eq.~\ref{formula}), leads to 
\begin{eqnarray}
&&L(\bm{\theta})\nonumber\\
&&= \frac{\beta^n \exp(-\beta \sum_{i=1}^n M_{1i})}{\sqrt{(2\pi)^n |\Sigma|}\sqrt{(2\pi)^n |\mK_n|}}\int_{\mM}\exp(\beta \bm{1}^\top \bm{x})\nonumber\\
&&\quad \times \int \exp\llp-\frac{1}{2}(\bm{\mu}-\bm{m}_1)^\top \Sigma_1^{-1}(\bm{\mu}-\bm{m}_1)\rrp \nonumber\\
&&\quad \times  \exp\llp-\frac{1}{2}(\bm{\mu}-\bm{m}_2)^\top \Sigma_2^{-1}(\bm{\mu}-\bm{m}_2)\rrp d\bm{\mu}d\bm{x}\nonumber\\
&&\begin{split}
&= \frac{\beta^n \exp(-\beta \sum_{i=1}^n M_{1i})}{(2\pi)^n\sqrt{ |\Sigma||\mK_n|}}\\
&\quad \times \int_{\mM}\exp(\beta \bm{1}^\top \bm{x})\sqrt{(2\pi)^n|\Sigma_c|} \exp(C) d\bm{x}
\end{split}.
\end{eqnarray}

Next, we computed $m_c$, $\Sigma_c$, and $C$ in the formula. It follows from the standard argument in linear algebra that 
\begin{eqnarray}
	\Sigma_c &=& (\mK_n^{-1}+\Sigma^{-1})^{-1}= \mK_n\tilde{\mK_n}^{-1}\Sigma,\\
	\bm{m}_c &=& \Sigma_c \{\Sigma^{-1}(x+\beta \Sigma \bm{1})+\mK_n^{-1}\bm{\mu}_{\prior}\},\\
	 C&=&
	 -\frac{1}{2}\bm{\mu}_{\prior}^\top \tilde{\mK}_n^{-1}\bm{\mu}_{\prior}-\frac{1}{2}(\bm{x}+\beta\Sigma\bm{1})^\top \tilde{\mK}_n^{-1}(\bm{x}+\beta\Sigma\bm{1})\nonumber\\
	 && +\bm{\mu}_{\prior}^\top  (\mK_n+\Sigma)^{-1}(\bm{x}+\beta\Sigma\bm{1}),
\end{eqnarray}
where $\tilde{\mK}_n=\mK_n+\Sigma$. Rearranging the integrand so that it becomes a quadratic form of $\bm{x}$, we have
\begin{eqnarray}
&&L(\bm{\theta})\nonumber\\
&&
\begin{split}
&= \frac{\beta^n}{\sqrt{(2\pi)^n |\tilde{\mK}|}}  \exp\bigg\{-\beta \sum_{i=1}^n (M_{1i}-\mu_{\prior, i})-\frac{\beta^2}{2} \bm{1}^\top (\Sigma -\mK_n) \bm{1}\bigg\}\\
& \quad\times \int_{\mM} \exp\llp -\frac{1}{2}(\bm{x}-\tilde{\bm{\mu}})^\top \tilde{\mK}_n^{-1}(\bm{x}-\tilde{\bm{\mu}})\rrp d\bm{x},
\end{split}
\end{eqnarray}
where $\tilde{\bm{\mu}}=\bm{\mu}_{\prior}+\beta  \mK_n \bm{1}$. This is the desired conclusion.

\vspace{2ex}
\noindent
\textbf{Proof of Eq.~\eqref{pred}.}
Let $t^*_1$ be any data point may not be in the dataset. What we need to compute is
\begin{eqnarray}
	&&p(\mu^*\mid t_1^*, \mD)\nonumber\\
	&&=\int p(\mu^*\mid t_1^*, \bm{\mu}, \bm{t}_1)p(\bm{\mu}\mid \bm{t}_1, \bm{M}_1) d\bm{\mu}\nonumber\\
	&&= \frac{\int p(\mu^*\mid t_1^*, \bm{\mu}, \bm{t}_1)p(\bm{M}_1\mid \bm{\mu})p(\bm{\mu}\mid \bm{t}_1)d\bm{\mu}}{\int p(\bm{M}_1\mid \bm{\mu})p(\bm{\mu}\mid \bm{t}_1)d\bm{\mu}}.
\end{eqnarray}
Here, the denominator of the predictive distribution is exactly the same as the marginal likelihood, which has already been computed. Hence, it remains to show that the numerator becomes
\begin{eqnarray}
\begin{split}
&\beta^n \exp\llp -\beta \sum_{i=1}^n(M_{1i}-\mu_{\prior,i})-\frac{1}{2}\beta^2\lp ns^2-\sum_{i,j}\mK_{i,j}\rp\rrp\\
&\quad \times  \int_{\mM}  \mathcal{N}(\mu^*;~D(\bm{x}), \tau^2)\mathcal{N}\lp\bm{x};~ \tilde{\bm{\mu}}_n, \tilde{\mK}_n\rp d\bm{x}.
\end{split}
\end{eqnarray}
It is computed by using the formula (eq.~\ref{formula}) with respect to $\bm{\mu}$, with some tedious calculus.

\vspace{2ex}
\noindent
\textbf{Proof of Eq.~\eqref{pred_detect}.}
It follows from the result of eq.~\eqref{pred} that 
\begin{eqnarray}
	&&P(\delta=1\mid M^*_1, t^*_1, \mD)\nonumber\\
	&&= \int P(\delta=1\mid M^*_1, \mu^*, \mD)  p(\mu^*\mid t^*_1, \mD)d\mu^* \nonumber\\
	&&=\int_{-\infty}^{\infty} \int_{-\infty}^M \mathcal{N}(z; \mu^*, s^2)\nonumber\\
	&&\quad \times \frac{ \int_{\mathcal{M}}\mathcal{N}\lp \mu^*;~D^*(\bm{x}),~(\tau^*)^2\rp  \mathcal{N}\lp \bm{x};~\tilde{\bm{\mu}}_n,~\tilde{K}_n\rp dx}{\int_{\mathcal{M}} \mathcal{N}\lp \bm{x};~\tilde{\bm{\mu}}_n,~\tilde{K}_n\rp d\bm{x}} d\mu^* \nonumber\\
	&&= \int_{-\infty}^M\frac{ \int_{\mathcal{M}}\mathcal{N}\lp z;~D^*(\bm{x}),~s^2+(\tau^*)^2\rp  \mathcal{N}\lp \bm{x};~\tilde{\bm{\mu}}_n,~\tilde{K}_n\rp d\bm{x}}{\int_{\mathcal{M}} \mathcal{N}\lp\bm{x};~\tilde{\bm{\mu}}_n,~\tilde{K}_n\rp d\bm{x}}  dz\nonumber\\
	&&= \bE_\trunc\llp \Psi \lp\frac{M^*_1-D^*(\bm{X})}{\sqrt{s^2+(\tau^*)^2}}\rp\rrp.
\end{eqnarray}

\label{lastpage}

\end{document}